\documentclass[a4paper]{simple}

\usepackage{amssymb} 
\usepackage{graphicx} 
\usepackage{amsmath} 
\usepackage{natbib} 
\usepackage{epsfig} 
\usepackage{color}

\newcommand{\appropto}{\mathrel{\vcenter{ \offinterlineskip\halign{\hfil$##$\cr \propto\cr\noalign{\kern2pt}\sim\cr\noalign{\kern-2pt}}}}}

\title{Leverage-induced systemic risk under Basle II and other credit risk policies}

\author{Sebastian Poledna$^1$, Stefan Thurner$^{1,2,3}$$^{\ast}$\thanks{$^\ast$Corresponding author. Email: stefan.thurner@meduniwien.ac.at \vspace{12pt}}, J. Doyne Farmer$^{4,2}$ and John Geanakoplos$^{5,2}$\\
\vspace{12pt} \normalfont{ $^1$Section for Science of Complex Systems, Medical University of Vienna, Spitalgasse 23, A-1090, Austria\\
$^2$Santa Fe Institute, 1399 Hyde Park Road, Santa Fe, NM 87501, USA\\
$^3$IIASA, Schlossplatz 1, A-2361 Laxenburg, Austria\\
$^4$Institute for New Economic Thinking at the Oxford Martin School and Mathematical Institute, 24-29 St. Giles, University of Oxford, Oxford OX1 3LB, UK\\
$^5$Department of Economics, Yale University, New Haven CT }\vspace{12pt} }

\begin{document}

\maketitle 
\begin{abstract}
	We use a simple agent based model of value investors in financial markets to test three credit regulation policies. The first is the unregulated case, which only imposes limits on maximum leverage. The second is Basle II and the third is a hypothetical alternative in which banks perfectly hedge all of their leverage-induced risk with options. When compared to the unregulated case both Basle II and the perfect hedge policy reduce the risk of default when leverage is low but increase it when leverage is high. This is because both regulation policies increase the amount of synchronized buying and selling needed to achieve deleveraging, which can destabilize the market. None of these policies are optimal for everyone: Risk neutral investors prefer the unregulated case with low maximum leverage, banks prefer the perfect hedge policy, and fund managers prefer the unregulated case with high maximum leverage. No one prefers Basle II. 
\end{abstract}

\newpage

\section{Introduction} \label{intro}

The recent crash in home and mortgage prices, and the ensuing global recession, has brought forth numerous proposals for the regulation of leverage. The trouble is that many of these proposals ignore the mechanism of the leverage cycle, and thus might unwittingly do more harm than good.

Leverage is defined as the ratio of assets held to wealth. A homeowner who buys a house for \$100 by putting down \$20 of cash and borrowing the rest is leveraged 5 to 1. One reason leverage is important is that it measures how sensitive the investor is to a change in asset prices. In the case of the homeowner, a \$1 or 1\% decline in the house price represents a 5\% loss in his wealth, since after he sells the house and repays the \$80 loan he will only have \$19 out of his original \$20 of capital. Limiting leverage therefore seems to protect investors from themselves, by limiting how much they can all lose from a 1\% fall in asset prices. Basle II effectively puts leverage limits on loans banks can give to investors, and furthermore it ties the leverage restriction to the volatility of asset prices: if asset prices become more likely to change by 2\% instead of 1\%, then Basle II curtails leverage even more. At first glance this seems like good common sense.

The leverage cycle, however, does not arise from a once and for all exogenous shock to asset prices, whose damages to investors can be limited by curtailing leverage. On the contrary, the leverage cycle is a process crucially depending on the heterogeneity of investors. Some investors are more optimistic than others, or more willing to leverage and buy than others. When the market is doing well these investors will do well and via their increased relative wealth and their superior adventurousness, a relatively small group of them will come to hold a disproportionate share of the assets. When the market is controlled by a smaller group of agents who are more homogeneous than the market as a whole, their commonality of outlook will tend to reduce the volatility of asset prices. But this will enable them, according to the Basle II rules, to leverage more, which will give them a still more disproportionate share of the assets, and reduce volatility still further. Despite the leverage restrictions intended from Basle II, the extremely low volatility still gives room for very high leverage.

At this point some exogenous bad luck that directly reduces asset prices will have a disproportionate effect on the wealth of the most adventurous buyers. Of course they will regard the situation as an even greater buying opportunity, but in order to maintain even their prior leverage levels they will be forced to sell instead of buying. At this point volatility will rise and the Basle II lending rules will force them to reduce leverage and sell more. The next class of buyers will also not be able to buy much because their access to leverage will also suddenly be curtailed. The assets will cascade down to a less and less willing group of buyers. In the end, the price of the assets will fall not so much because of the exogenous shock, but because the marginal buyer will be so different from what he had been before the shock. Thus we shall show that in some conditions, Basle II not only would fail to stop the leverage build up, but it would make the deleveraging crash much worse by curtailing all the willing buyers simultaneously. The policy itself creates systemic risk.

We shall also see that another apparently sensible regulation can lead to disaster. Common sense suggests it would be safer if the banks required funds to hedge their positions enough to guarantee they can pay their debts before they could get loans. The trouble with this idea is that when things are going well, the most adventurous leveragers will again grow, thereby lowering volatility. This lower volatility will reduce their hedging costs, and enable them to grow still faster and dominate the market, reducing volatility and hedging costs still more. Bad luck will then disproportionately reduce the wealth of the most enthusiastic buyers. But more importantly, it will increase volatility and thus hedging costs. This will force further selling by the most enthusiastic buyers, and limit the buying power of the next classes of potential owners. In just the same way as Basle II, the effort to impose common sense regulation of leverage can create bigger crashes.

In recent years a variety of studies including \cite{fostel08}, \cite{geanakoplos10}, \cite{adrian08}, \cite{brunnermeier09}, \cite{thurner09}, and \cite{caccioli12} have made it clear that deleveraging can cause systemic financial instabilities leading to market failure, as originally discussed by \cite{minsky92}. The specific problem is that regulatory action can cause synchronized selling, thereby amplifying or even creating large downward price movements. In order to stabilize markets a variety of new regulatory measures have been proposed to suppress such behavior. But do these measures really address the problem? 

As pointed out in \cite{thurner13} systemic risk is not only related to network properties but it is a multiplex network concept. By this we mean that systemic risk happens on various layers of the financial system, which can all influence each other. The networks involved include: borrowing and lending relationships (which can be further broken down into explicit contractual obligations, i.e. counterparty exposures, and implicit relationships, such as roll-over of overnight loans), insurance (derivative) contracts, collateral obligations, market impact of overlapping asset portfolios and network of cross-holdings (holding of securities or stocks of fellow banks). In this paper we focus on the systemic risk component of overlapping portfolios. Our example here is simple, as there is only one risky asset. Contagion is transmitted between agents when they buy or sell the asset, and as we will see, the use of leverage can lead to market crises. A key point in this study is that crises emerge endogenously, under normal operation of the model -- there are long periods where the market is relatively quiet, but due to the build up of leverage, the market becomes more sensitive to small fluctuations (which would at other times have negligible effect). 

Of particular interest here are leverage constraints, which are a significant part of financial regulation. These constraints are implemented in numerous ways, most influential in the form of capital adequacy rules in the Basle II framework and as margin requirements and debt limits in the Regulations T, U, and X of the Federal Reserve System. Margin requirements were established in the wake of the 1929 stock market crash with the belief that margin loans led to risky investments resulting in losses for lenders \citep{fortune00}. \citet{fortune01} discusses the regulation, historical background, accounting mechanics and economic principles of margin lending according to Regulations T, U, and X. 

In comparison to straightforward leverage constraints, the Basle II capital adequacy rules classify and weight assets of banks according to credit risk. Banks regulated under the Basle II framework are required to hold capital equal to $8\%$ of risk-weighted assets. A recent case study of the Bank of Canada discusses unweighted leverage constraints as a supplement to existing risk-weighted capital requirements \citep{bordeleau09}. The second of the Basle II Accords (Basle II) capital adequacy regulations added a significant amount of complexity and sophistication to the calculation of risk-weighted assets. In particular, banks are encouraged to use internal models, such as value-at-risk (VaR), to determine the value of risk-weighted assets according to internal estimations. In a nontechnical analyses of the Basle II rules, \citet{balin08} provides an easy accessible analysis of both the Basle I and Basle II framework.

\citet{lo12} provide an extensive overview of leverage constraints, pointing out that regulatory constraints on leverage are generally fixed limits that do not vary over time or with changing market conditions, and suggest that from a microprudential perspective fixed leverage constraints result in large variations in the level of risk. Recent studies of central banks also conclude that current regulatory leverage constraints are inadequate \citep{bhattacharya11,christensen11}. From a macroprudential perspective internal estimations of banks appear to be cyclically biased in determining the value of risk-weighted assets, contributing to a procyclical increase in global leverage \citep{bordeleau09}. \citet{danielsson01} and \citet{danielsson02} have also argued that the Basle II regulations fail to consider the endogenous component of risk, and that the internal models of banks can have destabilizing effects, inducing crashes that would otherwise not occur.

Computational agent-based models have gained popularity in economic modeling over the last decades and are able to reproduce some empirical features of financial markets that traditional approaches cannot replicate\citep{lebaron08}. An advantage of this approach is the ability to implement institutional features accurately and to be able to simulate any model setup, without the constraints of analytic tractability. An extensive review of financial multi-agent models can be found in \cite{lebaron06} and \cite{hommes06}. \cite{lebaron06} has focused more on models with many types of agents and \cite{hommes06} concentrating more on models with a few types of agents and the effects of heterogeneous strategies.

In this paper we use an agent-based financial market model introduced by \citet{thurner09} to test the performance of several credit regulation policies. The model introduced by \citet{thurner09} will be used as a baseline. In this work, the model is extended to allow short selling and to incorporate different regulation policies. The use of this simulation model allows us to explicitly implement and test any given regulatory policy. We test three different cases: (1) An unregulated market, (2) the Basle II framework and (3) a hypothetical regulatory policy in which banks completely hedge against possible losses from providing leverage (while charging their clients the hedging costs). We find that when leverage is high both of the regulatory schemes fail to guard against systemic financial instabilities, and in fact result in even higher rates of default than no regulation at all. The reason for this is that both regulatory policies compel investors to deleverage just when this is destabilizing, triggering failures when they would otherwise not occur.

Agent-based models have often been criticized for making arbitrary assumptions, particularly concerning agent decision making. We address this problem here by keeping the model simple and making a minimum of behavioral assumptions. There are four types of investors: 
\begin{enumerate}
	\item {\it Fund managers} are perfectly informed value investors that all see the same perfect valuation signal. They buy when the market is underpriced and sell when it is overpriced. Fund managers are risk-neutral. 
	\item {\it Noise traders} are inattentive value investors. They buy or sell when the market is under or over priced, but do so less efficiently than the fund managers. Noise traders are risk-neutral. 
	\item {\it Banks} loan money to the fund managers to allow them to leverage. Banks are risk-averse, though the extent to which this is true differs depending on the regulator scheme. 
	\item {\it Fund investors} place or withdraw money from the fund managers based on a historical average of each fund's performance. 
\end{enumerate}

The fund managers, which are the primary focus of this article, are heterogeneous agents, i.e. there are multiple funds with different levels of aggression. In contrast, the other three types are representative agents, i.e. there is a single noise trader, a single bank, and a single fund investor. The primary function of the noise traders is to provide a background price series to generate opportunities for the fund investor; the function of the fund investors is to guarantee that the price dynamics can be run in a steady state without the fund managers becoming infinitely wealthy. The bank is there to lend money to the funds, but it is a backdrop -- the bank is infinitely solvent. (It is nonetheless useful to examine welfare measures, such as the losses to the bank, rate of default on loans, etc.)

The postulated behaviors of each type of agent are reasonably generic, and in some cases, such as for the banks, we are merely codifying behaviors that are essentially mechanical and legally mandated by the terms of contractual agreement or regulation. Most importantly, our results are relatively insensitive to assumptions. The fact that this is a simulation model has the important advantage of allowing us to quantitatively and explicitly evaluate any regulatory scheme. This model is useful because it can incorporate more realism than a typical stylized equilibrium model \citep{bouchaud08,farmer09,lux09,thurner11}. This joins a growing class of agent-based models for testing economic policies that attempt quantifiable, reproducible and falsifiable results, whose parameters are -- at least in principle -- observable in reality. Recent studies that use agent-based models for testing economic policies include \citet{demary08}, \citet{haber08}, \citet{westerhoff08,westerhoff12}, \citet{kerbl10} and \citet{hermsen10}\nocite{lebaron08}.

In Section \ref{baselinemodel}, following \citet{thurner09} we present the model that serves as the baseline for comparison of regulatory schemes. In Section \ref{regulatorymeasures} the modifications necessary to implement the regulatory measures are explained. Section \ref{results} presents simulation results comparing both the Basle II-type regulation and the full perfect-hedge regulation schemes under various leverage levels in the financial system. In Section \ref{discussion} we conclude, discussing why Basle II-type systems -- even in their most ideal form -- destabilize the market just when stability is most needed, i.e. in times of high leverage levels in the system. At the heart of this problem are synchronization effects of financial agents in time of stress.

\section{The baseline model} \label{baselinemodel} 
\begin{figure}
	\begin{center}
		\includegraphics[width=13cm]{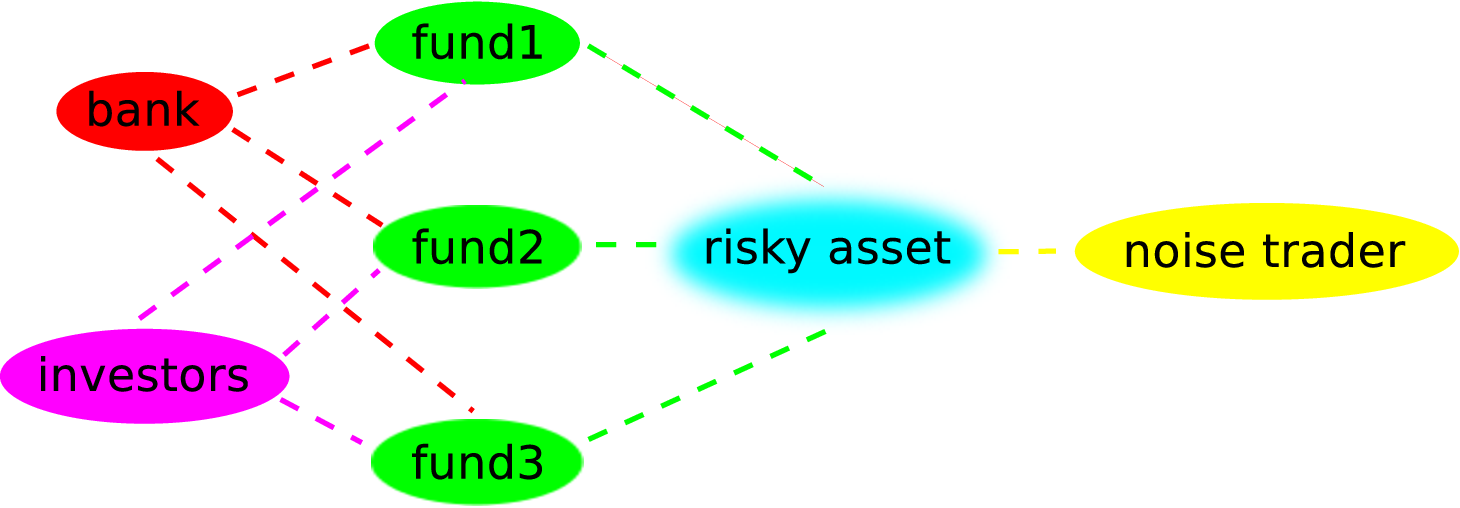} 
	\end{center}
	\caption[Schematic structure of the model]{Schematic structure of the model: Fund managers buy a single risky asset when it is underpriced and sell when it is overpriced. Noise traders also buy or sell when the market is under or over priced, but do so less efficiently than the fund managers -- the noise traders and fund managers have the same notion of value. A representative bank loans money to the fund managers to allow them to leverage. In the Basle II, and the perfect-hedge scheme fund managers pay fees for receiving loans, see Eqs. (\ref{fundWealthSpreadLong}) and (\ref{fundWealthSpreadShort}). Investors place or withdraw money from the funds based on their historical performance. Funds are heterogeneous based on the aggression of their investment strategy; more aggressive funds tend to use more leverage. For fund manager profits see figure \ref{fundimpact}(c). In the model no actual fees are paid. The plot corresponds to a hypothetical situation, which is certainly correct for small hedge funds. \label{network}} 
\end{figure}

The baseline model represents a market that is unregulated except for a maximum leverage requirement. It is an agent-based model with four different types of agents as described in \cite{thurner09}. There is only a single asset, without dividends and consumption, and investors are given a choice between holding the asset or holding cash. Prices are formed via market clearing. In the subsequent paragraphs we give an overview of each type of agent, and then in the remainder of this section we describe their behavior in more detail.

In figure \ref{network} we show a schematic structure of the model. The first type of agents are {\em fund managers}, e.g. hedge funds or proprietary trading groups. They are value investors who `buy low and sell high'. They use a strategy that translates a mispricing signal into taking a long position (buying a positive quantity of the asset) when the asset price $p(t)$ at time $t$ is below a perceived fundamental value $V$. We generalize the model of Thurner et al. by also allowing them them to take a short position when the asset is over-priced, i.e. one for $p(t)>V$. The demand of fund managers is denoted by $D_h(t)$, where the subscript $h$ refers to the fund manager. The fund managers are heterogeneous agents who differ in the aggressiveness with which they respond to buy and sell signals.

The second type of agent is a representative {\it noise trader}. This agent can be thought of as a weakly informed value investor, who has only a vague concept of the fundamental price, and thus buys and sells nearly at random, with just a small preference that makes the price weakly mean-revert around $V$. The demand of noise traders at time $t$ is $D_{n}(t)$. The noise trader is a representative agent, representing the pool of weakly-informed investors that cause prices to revert toward value.

The third type of agent is a {\em bank}. Fund managers can increase the size of their long positions by borrowing from the bank by using the asset as collateral. The bank limits lending so that the value of the loan is always (substantially) less than the current price of the assets held as collateral. This limit is called a minimum margin requirement. In case the asset value decreases so much that the minimum margin requirement is no longer sustained, the bank issues a margin call and the fund managers who are affected must sell assets to pay back their loans in order to maintain minimum margin requirements. This happens within a single timestep in the model. This kind of transaction is called margin trading and has the effect of amplifying any profit or loss from trading. If large price jumps occur and fund managers cannot repay the loan even by selling their complete portfolio, they default. In the baseline model, for simplicity interest rates for loans are fixed to zero and the bank sets a fixed minimum margin requirement, denoted by $\lambda_{\rm max}$.

The fourth type of agent is a representative {\em fund investor} who places or withdraws money from a fund according to performance. This agent should be viewed as a representative agent characterizing all investors, both private and institutional, who place money with funds that use leverage. The amount invested or redeemed depends on recent historical performance of each fund compared to a fixed benchmark return $ r_{b}$. Successful fund managers attract additional capital, unsuccessful ones lose capital. 

\subsection{Price formation} 

At each timestep $t$ asset prices $p(t)$ are formed via market clearing by equating the sum over the demand of the fund managers $D_{h}(t)$ and the noise traders $D_{\mathrm{n}}(t)$ to the fixed total supply $N$ of the asset, which represents the number of issued shares. The market clearing condition is 
\begin{equation}
	D_{\mathrm{n}}(t)+\sum_{h}D_{h}(t)=N. 
\end{equation}
\begin{figure}
	\begin{center}
		\includegraphics[width=6.5cm]{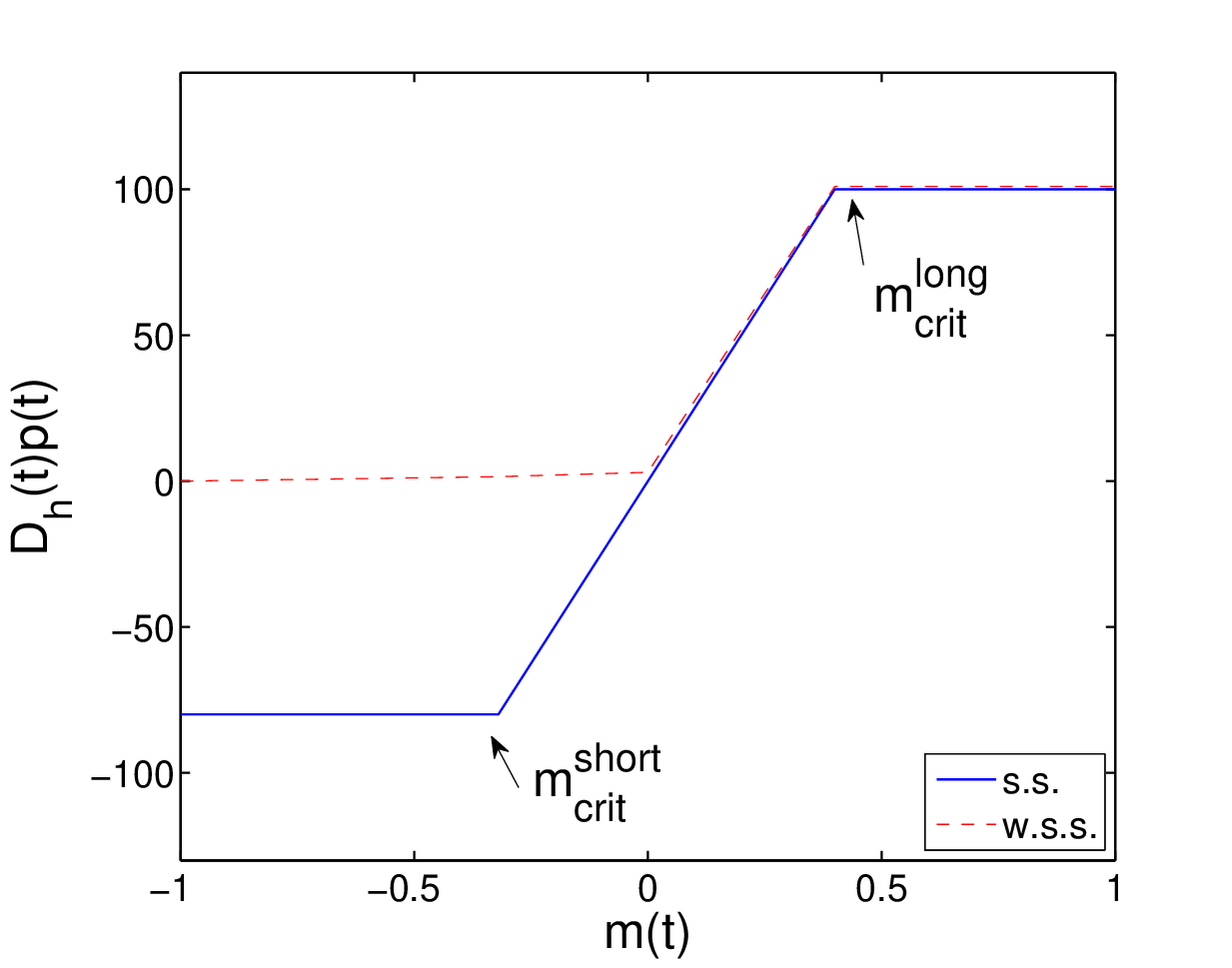} 
	\end{center}
	\caption[Demand function $D_{h}(t)p(t)$ of a fund manager as a function of the mispricing signal $m(t) = V - p(t)$.]{Demand function $D_{h}(t)p(t)$ of a fund manager as a function of the mispricing signal $m(t) = V - p(t)$. If the asset is underpriced the fund manager starts buying more and more assets as the price decreases, until the maximum margin requirement (leverage limit) is hit at $m = m_{\mathrm{crit}}$. Above this mispricing the demand remains flat. If short selling is banned (dashed line) the fund manager holds no assets if the asset is overpriced, whereas when short selling is allowed (solid line) the fund manager takes a negative position on the asset when it is overpriced. \label{valueInvestor}} 
\end{figure}

At every timestep the fund managers must decide how much of their total wealth $W_{h}(t)$ they are going to invest. The wealth of a fund manager is the sum of her cash position $M_{h}(t)$ and the current (dollar) value of the asset $D_{h}(t)p(t)$, 
\begin{equation}
	W_{h}(t)=D_{h}(t)p(t)+M_{h}(t). 
\end{equation}
When a fund manager borrows cash for leveraging positions, the cash $M_{h}(t)$ is negative, representing the fact that she has spent all her money and is in debt to the bank. Leverage $\lambda_{h}$ is defined as the ratio between the fund manager's portfolio value and her wealth, 
\begin{equation}
	\lambda_{h}(t)= \frac{D_{h}(t)p(t)}{W_{h}(t)}= \frac{D_{h}(t)p(t)}{D_{h}(t)p(t)+M_{h}(t)}. \label{leverageEQ} 
\end{equation}
In case of short selling leverage is defined as the ratio of the asset side in the balance sheet and the wealth, 
\begin{equation}
	\lambda_{h}(t)= \frac{W_{h}(t)-D_{h}(t)p(t)}{W_{h}(t)}= \frac{M_{h}(t)}{D_{h}(t)p(t)+M_{h}(t)}. \label{leverageEQshort} 
\end{equation}

The fund managers are value investors who base their demand $D_{h}(t)$ on a {\it mispricing signal}, $m(t)=V-p(t)$. For simplicity the perceived fundamental value $V$ is fixed at a constant value, which is the same for all fund managers and noise traders. Figure~\ref{valueInvestor} shows demand $D_{h}(t)$ for a fund manager $h$ as a function of the perceived mispricing. As the mispricing increases, the fund manager wants a linear increase of the value of the portfolio, $D_{h}(t)p(t)$. However, this is bounded when it reaches the maximum leverage level $\lambda_{\mathrm{max}}$, set by the bank. Fund managers differ in their {\em aggression parameter} $\beta_{h}$, which quantifies how strongly they respond to the mispricing signal $m(t)$. The fund manager's demand function, $D_{h}(t)=D_{h}(t,p(t))$, can be written as 
\begin{equation}
	\label{fundDemand} D_{h}(t) = 
	\begin{cases}
		0 &\text{if }m(t)<0\\
		\lambda_{\mathrm{max}}W_{h}(t)/p(t) &\text{if }m(t)> m_{\mathrm{crit}}\\
		\beta_{h}m(t)W_{h}(t)/p(t) &\text{otherwise}, 
	\end{cases}
\end{equation}
and when short selling is allowed\footnote{See \citet{kerbl10}: p. 6 for fund manager's demand when short selling is allowed.}, 
\begin{equation}
	D_{h}(t) = 
	\begin{cases}
		(1-\lambda_{\mathrm{max}})W_{h}(t)/p(t) &\text{if }m(t) \leq m_{\mathrm{crit}}^{\mathrm{short}}\\
		\lambda_{\mathrm{max}}W_{h}(t)/p(t) &\text{if }m(t)> m_{\mathrm{crit}}^{\mathrm{long}}\\
		\beta_{h}m(t)W_{h}(t)/p(t) &\text{otherwise}. 
	\end{cases}
	\label{fundDemandshort} 
\end{equation}
The parameter $m_\mathrm{crit} > 0$ is the critical value of the mispricing signal above which the fund manager is forced to flatten demand when short selling is not allowed, and similarly $m_{\mathrm{crit}}^{\mathrm{short}} < 0$ and $m_{\mathrm{crit}}^{\mathrm{long}} > 0$ are the corresponding parameters when short selling is allowed. Notice that $\lambda_{\mathrm{max}} > 1$ is strictly required for short-selling.

To summarize, consider the case where short selling is allowed. When the asset is strongly overpriced the value of the fund manager's position is constant at the value $D_h(t) p(t) = (1-\lambda_{\mathrm{max}})W_{h}(t)$. Similarly when the asset is strongly underpriced the value is constant at $\lambda_{\mathrm{max}}W_{h}(t)$. Anywhere in between the value is $\beta_{h}m(t)W_{h}(t)$, i.e. it is proportional to the aggression parameter $\beta_{h}$, the mispricing $m(t)$, the wealth $W_{h}(t)$.

Every fund manager is required by the bank to maintain $\lambda_{h}(t)\leq\lambda_{\mathrm{max}}$. If this condition would be violated the fund manager must adjust her demand to buy or sell assets to ensure that such a violation does not happen. This is known as meeting a \emph{margin call}. The cause of a margin call can be either because the price drops from $p(t-1)$ to $p(t)$, causing $W_{h}(t)$ to fall by a larger percentage than the asset price (because of leverage), or because the wealth drops from $W_{h}(t-1)$ to $W_{h}(t)$ due to withdrawals (redemptions) from investors, as will be discussed below. Fund managers adjust their positions within each timestep to make sure that this condition is never violated.

The \textit{noise trader} demand is formulated in terms of the value $\xi_{\mathrm{n}}(t)= D_{\mathrm{n}}(t)p(t)$, whose logarithm follows an Ornstein--Uhlenbeck random process of the form 
\begin{equation}
	\log\xi_{\mathrm{n}}(t)=\rho\log\xi_{\mathrm{n}}(t-1)+\sigma_{\mathrm{n}}\chi (t)+(1-\rho)\log(VN), 
\end{equation}
where $\chi$ is independent and normally distributed with mean zero and standard deviation one, and where $0 < \rho<1$. The Ornstein–-Uhlenbeck process is a widely used approach to model currency exchange rates, interest rates, and commodity prices stochastically. In the limit as $\rho\rightarrow1$, i.e. without the mean reversion, the $\log$-returns $r(t)=\log p(t+1)-\log p(t)$ of the asset price are normally distributed. We choose a value of $\rho$ close to one so that without fund managers the price is weakly mean reverting around the fundamental value $V$ and the $\log$-returns are nearly normally distributed.

\subsection{Defaults}

If the fund manager's wealth ever becomes negative, i.e. if $W_{h}(t)<0$, the fund manager defaults and goes out of business. The fund manager must then sell all assets ($D_{h}(t)=0$) and use the revenue to pay off as much of the loan as possible. All remaining loss is born by the bank, causing capital shortfall, which has to be provided by the government or a bailout fund. For simplicity we assume banks always receive the necessary bailout funds and they continue to lend to other fund managers as before. $T_{\rm re-intro}$ timesteps later the defaulting fund manager re-emerges as a new fund manager, as described below.

\subsection{Wealth dynamics of the fund managers} 

Initially each fund manager has the same endowment, ${W}_{h}(0)=W_0$. The wealth of the fund manager then evolves according to 
\begin{equation}
	\label{fundWealth} W_{h}(t)=W_{h}(t-1)+D_{h}(t-1)[p(t)-p(t-1)]+F_{h}(t), 
\end{equation}
where $D_{h}(t-1)[p(t)-p(t-1)]$ reflects the profits and losses from trading of the fund manager's portfolio and $F_{h}(t)$ quantifies the deposits or withdrawals of the fund investor. 

The amount of cash a fund manager would have if selling all assets at the current price is 
\begin{equation}
	\label{fundCapital} \tilde{M_h}(t)=D(t-1)p(t)+M(t-1). 
\end{equation}

The fund investor deposits or withdraws from each fund manager based on a moving average of the recent performance which is measured by the rate of return. This prevents the wealth of the fund managers from growing indefinitely, and makes possible well defined statistical averages for properties such as returns and volatility. Let 
\begin{equation}
	r_{h}(t) = \frac{D_{h}(t-1)[p(t)-p(t-1)]}{W_{h}(t-1)} , 
\end{equation}
be the rate of return of fund manager $h$. The fund investors base their decisions on the recent performance of fund $h$, measured as an exponential moving average of the rate of return 
\begin{equation}
	r_{h}^{\mathrm{perf}}(t)=(1-a)\,r_{h}^{\mathrm{perf}}(t-1)+a\,r_{h}(t). 
\end{equation}
The flow of capital in or out of the fund manager is given by 
\begin{equation}
	\label{fundFlow} F_{h}(t)=\max\left[-1,b \left(r_{h}^{\mathrm{perf}}(t)- r_{b}\right)\right] \max\left[0,\tilde{M_h}(t)\right], 
\end{equation}
where $b$ is a parameter controlling the fraction of capital withdrawn or invested, $a$ is the moving average parameter, and $ r_{b}$ is the benchmark return of the investors. This process is well documented, see e.g. \citet{Busse01,Chevalier97,Delguercio02,Remolona97,Sirri98}. Fund investors cannot take out more cash than the fund manager has.

In case a fund manager's wealth falls below a critical threshold, $W_{\rm crit}$, the fund manager goes out of business. This avoids the possibility of ``zombie funds'', which persist for many timesteps with nearly no wealth and no relevance for the market. After $T_{\rm re-intro}$ timesteps, funds that default are replaced with a new fund with initial wealth $W_0$ and the same aggression parameter $\beta_{h}$.

\section{Implementation of regulatory measures} \label{regulatorymeasures} 

We now introduce two regulatory policies. The first is the regulatory measure to reduce credit risk encouraged by the Basle II framework. The second regulation policy is an alternative proposal in which all risk associated with leverage is required to be perfectly hedged by options.

\subsection{Basle II} The Basle II scheme that we employ here models the risk control policies used by banks regulated under the Basle II framework. These banks use internal models to determine the value of their risk-weighted assets. The banks are allowed discretion in their risk control as long as it is Basle II compliant. We implement a Basel compliant model that is used in practice by banks. 

\subsubsection{Credit exposure} According to the Basle II capital adequacy rules banks have to allocate capital for their credit exposure. The capital can be reduced through credit mitigation techniques, such as taking collateral (e.g. securities or cash) from the counterparty on the loan. This reduces their net adjusted exposure. However, to provide a safety margin haircuts are also applied to both the exposure and the collateral. {\it Haircuts} are percentages that are either added or subtracted depending on the context in order to provide a safety buffer. In the case of a collateralized loan, in the absence of haircuts, the net exposure taking the risk mitigating effect of the collateral into account is $E^* = E - k$, where $k$ is the value of the collateral. However, when haircuts are applied the value of the raw exposure $E$ is adjusted upward to $E(1 + H_e)$, where $H_e \ge 0$ is the haircut on the exposure, and the value of the collateral $k$ is adjusted downward to $k(1 - H_{col})$, where $H_{col} \ge 0$ is the haircut on the collateral. Thus the net exposure taking the risk mitigating effect of the collateral into account is\footnote{See \citet{bcbs06}: \S 147.} 
\begin{equation}
	\label{exposure} E^*=\max[0,E(1 + H^{\mathrm{}}_e)-k(1-H^{\mathrm{}}_{\mathrm{col}})], 
\end{equation}
The `max' is present to make sure the exposure is never negative.

\subsubsection{Haircuts} The basic idea is that undervaluing the asset creates a safety buffer which decreases the likelihood that at some point in the future the value of the collateral will be insufficient to cover possible losses. Increasing collateral has the dual effect of decreasing the chance that the collateral might be insufficient to cover future losses, and of decreasing the size of loans when collateral is in short supply. In general the size of the haircut, and hence the size of the exposure and collateral, depends on volatility.

Haircuts can either be standard supervisory haircuts, issued by regulatory bodies\footnote{See \citet{bcbs06}: \S 147 for recommendation on supervisory haircuts, e.g. haircut for equities listed on a recognized exchange is 25\%.}, or internal estimates of banks. Permission to use internal estimates is conditional on meeting a set of minimum standards\footnote{See \citet{bcbs06}: \S 156-165 for qualitative and quantitative standards.}. Here we use a simple approach -- used in practice by banks for their own internal estimates satisfying the Basle II standards -- and compute the haircuts $H^{\mathrm{}}_{\mathrm{col}}$ using a formula that sets a minimum floor on the haircut $H_{\mathrm{min}}$ with an additional term that increases with the historical volatility $\sigma(t)$, measured in terms of standard deviation of log-returns over $\tau$ time steps. We assume a loan of time duration $T$ and a fixed cost $c$. The haircut is given by the formula 
\begin{equation}
	\label{haircut} H^{\mathrm{}}_{\mathrm{col}}(t)=\min\left[\max\left(H^{\mathrm{}}_{\mathrm{min}},\Phi \sigma(t) \sqrt{T}+c\right),1\right], 
\end{equation}
where $\Phi$ is a confidence interval set by the bank in accordance with the regulatory body\footnote{

According to the Basel framework $\Phi$ must be chosen such that given a historical volatility $\sigma(t)$ the haircut is sufficient in 99\% of cases. See \cite{bcbs06}: \S 156.}.

The use of the `min' and `max' guarantees that the haircut is always in the range $H_{\mathrm{min}} \le H_{\mathrm{col}} \le 1$.

The choice of a haircut implies a variable maximum leverage $\lambda^{\mathrm{adapt}}_{\mathrm{max}}(t)$. Suppose a bank makes a loan of size $L$ to a fund that pledges the shares of the asset it buys as collateral. Then the raw exposure is $E = L$ and the value of the shares of the asset is $k$. By definition the leverage is 
\begin{equation}
	\label{leverageEq2} \lambda^{\mathrm{adapt}}_{\mathrm{max}}(t) = \frac{k}{k - E}. 
\end{equation}
Since the exposure is cash no haircut is required and $H_e = 0$, but the collateral is risky, so $H_{\mathrm{col}} > 0$. If the net risk mitigated exposure $E^* = 0$, then combining Eqs.~(\ref{exposure}) and (\ref{leverageEq2}) implies 
\begin{equation}
	\label{lev_hair} \lambda^{\mathrm{adapt}}_{\mathrm{max}}(t) = \frac{1}{H_{\mathrm{col}}(t)}. 
\end{equation}
For short selling the fund borrows the asset (which is risky) and gives cash as collateral, so the situation is reversed, and $H_{\mathrm{col}} = 0$ with $H_e > 0$. In this case $H_e$ is set according to Eq.~(\ref{haircut}) with $H_e$ on the lefthand side instead of $H_{\mathrm{col}}$, and through similar logic 
\begin{equation}
	\lambda^{\mathrm{adapt}}_{\mathrm{max}}(t) = \frac{1}{H_e(t)}. 
\end{equation}
This shows explicitly how the haircuts impose a limit on the maximum leverage. For example if the haircut is $0.5$ the maximum leverage is $2$, whereas if the haircut is $0.1$ the maximum leverage is $10$.

To make correspondence with the unregulated case we set $H_{\mathrm{min}} = \frac{1}{\lambda_{\rm max}}$ and $\Phi = \frac{1}{\lambda_{\rm max}\sigma_{b}}$, where $\sigma_{b}$ is a benchmark volatility that serves as an exogenous parameter. Furthermore, for simplicity we set transaction costs to zero, i.e. $c=0$, and the holding duration of the collateral to one timestep, $T=1$. Then by combining Eqs. (\ref{haircut}) and (\ref{lev_hair}) we see that the adaptive maximum leverage explicitly depends on volatility, 
\begin{equation}
	\label{lambdamaxbasel} \lambda^{\mathrm{adapt}}_{\mathrm{max}}(t)= \max\left[ \lambda_{\mathrm{max}} \min\left(1,\frac{\sigma_{b}}{\sigma(t)}\right),1\right]. 
\end{equation}
If the historical volatility $\sigma(t)\leq \sigma_{b}$ then full maximum leverage $\lambda_{\mathrm{max}}$ can be used by the fund managers. When the simulation is operating under the Basle II scheme the $\lambda_{\mathrm{max}}$ in the fund manager's demand of equation \eqref{fundDemand} is replaced by $\lambda^{\mathrm{adapt}}_{\mathrm{max}}(t)$ in equation (\ref{lambdamaxbasel}). 

\subsubsection{Spreads} \label{spreads}

To determine interest rates on loans to fund managers, banks add a risk premium (spread) $S$ to a benchmark interest rate $i_{b}$. Usually $S$ is determined by rating a customer, but in many cases such as margin trading a fixed risk premium is used for all customers of a given type. We use a fixed spread $S$ for all fund managers. The interest rate for the fund manager $h$ is 
\begin{equation}
	\label{fundInterestSpread} i_{h} = i_{b}+S. 
\end{equation}
To implement borrowing costs in the agent-based model we set the benchmark interest rate $i_{b}=0$ and add a term accounting for the spread $S$ to equation \eqref{fundWealth}. Fund managers always pay the borrowing costs for the previous timestep. In case the fund manager takes a leveraged long position ($M_h$ is negative) her wealth is 
\begin{equation}
	\label{fundWealthSpreadLong} W_{h}(t)=W_{h}(t-1)+D_{h}(t-1)[p(t)-p(t-1)]+F_{h}(t)+M_{h}(t-1)S, 
\end{equation}
and in case of short selling, where the demand is negative, 
\begin{equation}
	\label{fundWealthSpreadShort} W_{h}(t)=W_{h}(t-1)+D_{h}(t-1)[p(t)-p(t-1)]+F_{h}(t)+D_{h}(t-1)p(t-1) S. 
\end{equation}
The maximum amount the fund investor can redeem from the fund manager has to be adjusted from equation \eqref{fundCapital} to 
\begin{eqnarray}
	\label{fundCapitalSpreadLong} \tilde{M_h}(t)&=&D(t-1)p(t)+M(t-1)[1+S]\qquad \qquad \qquad \qquad \,\,\, [{\rm long}] \nonumber \\
	\tilde{M_h}(t)&=&D(t-1)p(t)+M(t-1)+D_{h}(t-1)p(t-1)S \qquad [{\rm short}] . 
\end{eqnarray}
This guarantees that obligations to banks are satisfied {\em before} those of investors, as is the usual practice.

\subsection{The perfect-hedge scheme} 

The idea behind the perfect-hedge scheme is that, in addition to holding the shares of the asset as collateral, banks require all loans to be hedged by options. Thus barring default of the issuer of the option, the loan is completely secure. 

\subsubsection{Hedging} 

We first consider the case where the fund is long, in which case the bank requires it to buy a put with strike price $K_{\mathrm{put}}$. For simplicity we treat the loan as an overnight loan and thus the put has a maturity of one day. To make sure that the loan can be repaid, the value of the asset must equal the value of the loan. If the price of the asset when the loan is made is $p(t)$, from Eq.~(\ref{leverageEq2}) the fraction by which it can drop in price before the value of the collateral is less than that of the original loan is $(k - E)/E = 1/\lambda_h(t)$. The strike price is thus 
\begin{equation}
	K_{\mathrm{put}}(t) =p(t) \left(1-\frac{1}{\lambda_{h}(t)} \right), 
\end{equation}
and via similar reasoning, in the case where the fund is short it buys a call option with strike price 
\begin{equation}
	K_{\mathrm{call}}(t) =p(t) \left(1+\frac{1}{\lambda_{h}(t)-1}\right). 
\end{equation}

Option prices are assumed to obey the Black-Scholes formula \citep{black73}. The option price is calculated based the following parameters: The spot price of the underlying asset is $p(t)$, the risk-free interest rate $r = 0$, and the volatility $\sigma = \theta \sigma(t)$, where $\sigma(t)$ is the historical volatility\footnote{We use the historical volatility $\sigma(t)$, defined as the standard deviation of the $\log$-returns of the underlying asset over $\tau$ timesteps. Volatility is multiplied by an arbitrary factor $\theta$ to gauge the length of one timestep. In the simulations we set $\theta=5$}. When the fund is long the strike price $P_h$ of the put option is $K=K_{\mathrm{put}}$ and when it is short the strike price $C_h$ of the call option is $K_{\mathrm{call}}$. Note that the option prices depend on the leverage through the strike prices; as the leverage increases the options get closer to be in the money and so their price increases.

To implement the hedging costs in the model we add a term for the option costs to equation \eqref{fundWealth}. In case the fund manager holds a long position hedging cost is $P_h(t) D_h(t-1)$ and the wealth becomes 
\begin{equation}
	\label{fundWealthHedgeLong} W_{h}(t)=W_{h}(t-1)+D_{h}(t-1)[p(t)-p(t-1)]+F_{h}(t)-D_{h}(t-1)P_{h}(t-1), 
\end{equation}
and similarly for short selling 
\begin{equation}
	\label{fundWealthHedgeShort} W_{h}(t)=W_{h}(t-1)+D_{h}(t-1)[p(t)-p(t-1)]+F_{h}(t)+D_{h}(t-1)C_{h}(t-1). 
\end{equation}
The maximum redemption of fund investors is adjusted for long positions 
\begin{equation}
	\label{fundCapitalHedgeLong} \tilde{M_h}(t)=D(t-1)p(t)+M(t-1))-D_{h}(t-1)P_{h}(t-1), 
\end{equation}
and for short selling is 
\begin{equation}
	\label{fundCapitalHedgeShort} \tilde{M_h}(t)=D(t-1)p(t)+M(t-1))+D_{h}(t-1)C_{h}(t-1). 
\end{equation}
Note that the only risk now for the bank is that in case the fund defaults, it will not be able to pay for the hedging costs $P_{h}(t-1)$ at the previous timestep. 

To make a correspondence to the spreads defined under the Basle II agreement, the {\em effective spreads} are\footnote{

To see this, note that $D_{h}(t)P_{h}(t)$ and $D_{h}(t)C_{h}(t)$ is the effective interest the funds have to pay for being long or short, respectively. To calculate an \emph{interest rate} we divide the interest by the loan size, i.e. 
\begin{equation}
	\frac{D_{h}(t)P_{h}(t)}{D_{h}(t)p(t)-W_h(t)}=\frac{D_{h}(t)P_{h}(t)}{-M_h(t)}. 
\end{equation}
Note that $M_h(t)$ is negative because the fund {\em owes} cash to the bank. By using $W_h(t)=D_{h}(t)p(t)/\lambda_h(t)$ in $D_{h}(t)P_{h}(t)/(D_{h}(t)p(t)-W_h(t))$, we obtain Eq. (\ref{fundInterestHedgeLong}). In case of short selling interest divided by the loan is $(D_{h}(t)C_{h}(t)) / (D_{h}(t)p(t))=C_{h}(t)/ p_h(t)$, as is Eq. (\ref{fundInterestHedgeLong}).} 
\begin{equation}
	\label{fundInterestHedgeLong} S_{h}^{[\rm long]}(t) = \frac{P_{h}(t)}{p(t){\left(1-\frac{1}{\lambda_{h}(t)}\right)}}\qquad {\rm and} \qquad S_{h}^{[\rm short]}(t) = \frac{C_{h}(t)}{p(t)}. 
\end{equation}

\subsubsection{Limit on hedging costs and maximum leverage} \label{limitsonhedging}

Absent any other constraints, under the above scheme there is the possibility that the maximum leverage could become arbitrarily large, and consequently the hedging costs could become very high. We prevent this by limiting leverage. To compute the limit on leverage needed to keep the hedging cost below a given threshold, for long positions we impose a maximum hedging cost $P_{\mathrm{max}}$. The corresponding dynamic maximum leverage $\lambda_{\mathrm{max}}^{\mathrm{hedge}}(t)$ can then be found by solving 
\begin{equation}
	P(p(t),\sigma(t),\lambda_{\mathrm{max}}^{\mathrm{hedge}}(t))=P_{\mathrm{max}}(t), 
\end{equation}
for $\lambda_{\mathrm{max}}^{\mathrm{hedge}}(t)$.

Similarly, just as in the Basle II scheme, there is the possibility that historical volatility (Eq. (\ref{fundInterestHedgeLong})) could become arbitrarily low. To prevent this we solve the same equation assuming a fixed volatility floor $\sigma_b$, which gives the maximum leverage as the solution of 
\begin{equation}
	P(p(t),\sigma_{b},\lambda_{\mathrm{max}}) = P_{\mathrm{max}}(t), 
\end{equation}
for $\lambda_{\mathrm{max}}$.

\begin{figure}
	\begin{center}
		\includegraphics[width=10.0cm]{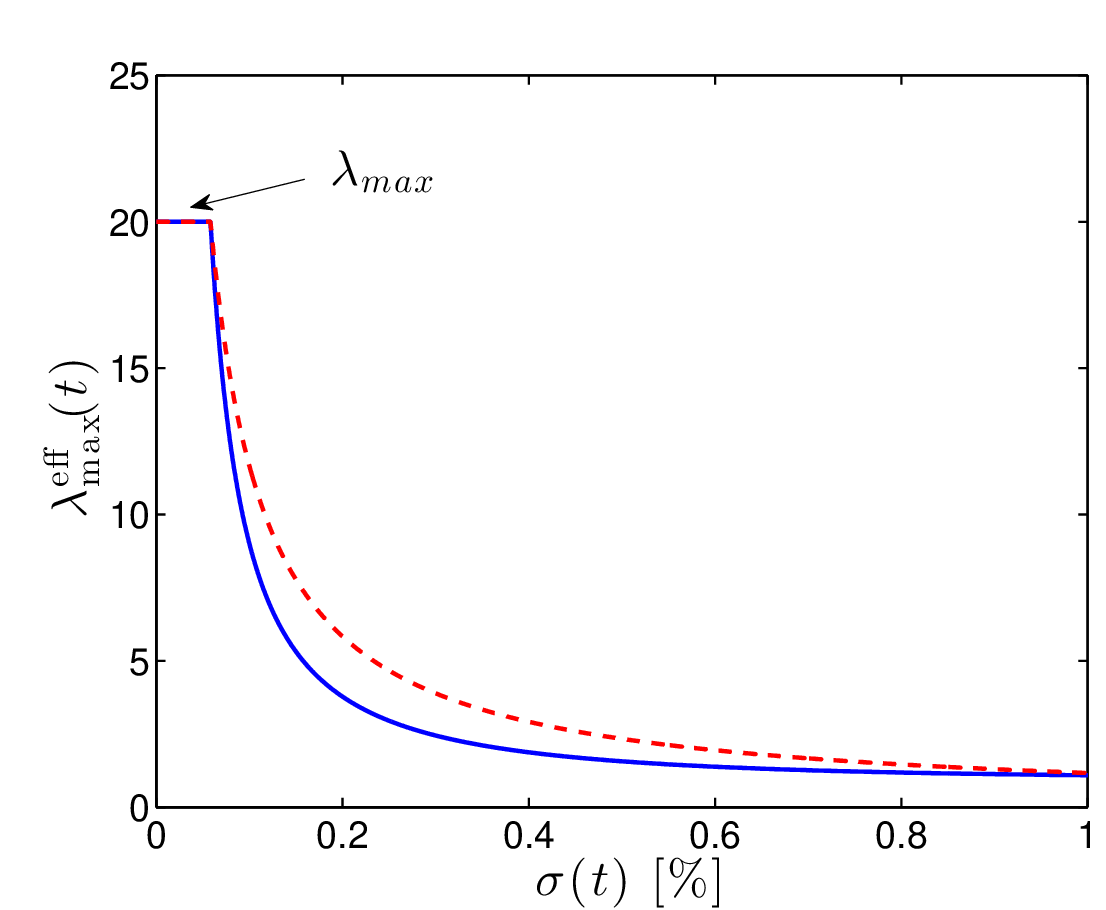} 
	\end{center}
	\caption[Maximum leverage function $\lambda_{\rm max}(t)$ for all fund managers as a function of historical volatility $\sigma(t)$.]{Maximum leverage function $\lambda^{\mathrm{adapt}}_{\mathrm{max}}(t)$ for fund managers as a function of historical volatility $\sigma(t)$ for the two regulation schemes. The dashed red curve shows the Basle II scheme of equation \eqref{lambdamaxbasel}, the solid blue curve shows the perfect-hedge scheme of equation \eqref{lambdamaxhedge}.} \label{lambdamax} 
\end{figure}

Finally, the leverage maximum $\lambda^{\mathrm{adapt}}_{\mathrm{max}}(t)$ is their minimum, i.e. 
\begin{equation}
	\label{lambdamaxhedge} \lambda^{\mathrm{adapt}}_{\mathrm{max}}(t)=\min\left[\lambda_{\mathrm{max}},\lambda_{\mathrm{max}}^{\mathrm{hedge}}(t)\right]. 
\end{equation}
The perfect-hedge version of the agent-based model is obtained by replacing the maximum leverage of equation \eqref{fundDemandshort} by equation \eqref{lambdamaxhedge}. Analogous equations hold for the short selling scenario. See figure \ref{lambdamax} for the dependence of $\lambda^{\mathrm{adapt}}_{\mathrm{max}}(t)$ as a function of $\sigma(t)$. 

\section{Results} \label{results} 

\subsection{Performance and efficiency indicators} \label{indicators} 

As performance indicators we use return to fund investors, profits to the fund managers, and probability of default. Since the fund investor actively invests and withdraws money from fund managers and funds have to cover for expenses, i.e. interest payments, the rate of return $r_{h}(t)$ does not properly capture the actual return to fund investors. To solve this accounting problem we compute the adjusted rate of return $r^{adj}$ to fund investors for any given period from $t=0$ to $t=T$. This is done by adjusting the wealth $W_{h}(t)$ of fund managers by the net flow of capital and the expenses within a given period, according to 
\begin{equation}
	\label{irr} r^{adj}_h(T) = \frac{W^{adj}_h(T)}{W_h(0)} - 1 
\end{equation}
with 
\begin{equation}
	W^{adj}_h(T) = W_h(T) - \sum^{T}_{i = 0}F_h(i) + f(T), 
\end{equation}
where $f$ is an adjustment for the expenses of the fund manager in a given period. In particular we have for long position under the Basle II scheme 
\begin{equation}
	f(T) = \sum^{T}_{i = 0}M_{h}(i)S, 
\end{equation}
and in the perfect hedge case 
\begin{equation}
	f(T) = \sum^{T}_{i = 0}D_{h}(i)P_{h}(i). 
\end{equation}
Expenses of fund managers for short positions are calculated in a similar way as described in section \ref{spreads} and \ref{limitsonhedging}. If a fund manager is out of business at $t=T$, $W_h(T)$ is set to zero and $r^{adj}_h(T)$ will be close to $-1$, deviations coming from the net flow of capital and expenses of the fund manager in the given period.

For management fees we use a hypothetical 2\% fixed fee for assets under management and a 20\% performance fee, paid by the fund investor to the fund managers. These fees are hypothetical and are not used as actual transactions in the model. They are only to indicate the profitability of fund managers under various conditions. If a fund manager goes out of business management fees are not paid.

We also consider the average annualized cost of capital, which is the effective interest rate $i_h(t)$ from equation \eqref{fundInterestSpread} and \eqref{fundInterestHedgeLong}. By annualized we mean that one simulation timestep represents five trading days and one year has 250 trading days. As performance indicators we monitor the standard deviation of the $\log$-returns $r(t)$ for all timesteps from an entire simulation run as an asset volatility `index'. Additionally we calculated the distortion as defined in \cite{westerhoff12}, i.e. the average absolute distance between log price and log fundamental $V$. Note that in our case the distortion is closely related to volatility. We explicitly checked that the distortion closely follows the curves of figure \ref{marketimpact}(a). As a measure for trading volume we use the average number of shares traded per timestep $1/HT\sum^H_{h=1}\sum^T_{t=1}\left|D_h(t) - D_h(t-1)\right|$, with $T$ denoting the number of timesteps in a simulation run and the number of fund managers $H = 10$. Finally, we measure the capital shortfall of banks by just keeping track of the amount of money they lose when funds default.

\subsection{Model Calibration: Choice of Parameters} \label{parameters} For all simulations we used 10 fund managers with $\beta_{h} = 5, 10, \ldots,50$, and simulation parameters $\rho=0.99$, $\sigma_{\mathrm{n}}=0.035$, $V=1$, $N=1\times10^9$, $r_b=0.003$, $a=0.1$, $b=0.15$, $W_0=2\times10^6$, $W_{\rm crit}=2\times10^5$, $T_{\rm re-intro}=100$, $\tau=10$, $\theta=5$, $\sigma_{b}=0.01175$ and $S=0.00015$. For most runs we used a range of $\lambda_{\mathrm{max}} \in [1,...,20]$. 

To test the robustness of the model we tested several different parameter sets, varied key parameters and motivated parameter choices in their economic context. A complete analysis of the robustness of the model is not feasible because of the large number of parameters, and because several parameters cannot be set independently.

Reducing or increasing the number of fund managers $H$ affects the wealth $W_h(t)$ of individual fund managers. With more fund managers individual banks accumulate less wealth. Therefore individual bankruptcies of fund managers become less severe resulting in lower capital shortfall for banks. Increasing the aggressiveness of fund managers $\beta_h$ causes them to react more to mispricing, resulting in more margin calls and subsequent defaults.

The parameters $\rho$, $\sigma_{\mathrm{n}}$, $V$, $N$ determine the demand of noise traders. $\sigma_{\mathrm{n}}$ is set to reflect typical stock price fluctuations. $N$, $V$ and $W_0$ are chosen to guarantee that fund managers have a low market share when they are introduced in market. We choose to arbitrarily set $V=1$ and choose $N$ and $W_0$ accordingly. The setting of $\rho\sim 1$ ensures that the deviation from the normal distribution is minimal. With $\rho=0.99$ the typical fluctuation in volatility is about $1\%$, which is a reasonable volume given real stock values.

The benchmark return $r^{\mathrm{b}}$ plays the important role of determining the relative size of hedge funds vs. noise traders. If the benchmark return is set very low then funds will become very wealthy and will buy a large quantity of the asset under even small mispricings, preventing the mispricing from ever growing large. This effectively induces a hard floor on prices. If the benchmark return is set very high, funds accumulate little wealth and play a negligible role in price formation. The interesting behavior is observed at intermediate values of $r^{\mathrm{b}}$ where the funds' demand is comparable to that of the noise traders. The parameter "$a$ " governs the exponential moving average of the performance of fund managers and the parameter "$b$" controls the sensitivity of withdrawals or contributions of fund investors. Both parameters $a$ and $b$ are set empirically, following work from \cite{Busse01},\cite{Chevalier97}, \cite{Delguercio02}, \cite{Remolona97} and \cite{Sirri98}.

Using a positive survival threshold for removing funds $W_{\rm crit}$ avoids the creation of “zombie funds” that persist for long periods of time with almost no wealth. Setting $T_{\rm re-intro}=100$, which controls the reintroduction of funds after bankruptcies, corresponds to 2 years under the calibration that one timestep is five days, which we think is a reasonable value. Higher settings for $T_{\rm re-intro}$ result in simulations with few fund managers present at turbulent times.

The parameter $\tau$ is chosen to measure historical volatility over a short window. Choosing different windows $\tau$ $5 < \tau < 20$ does not substantially influence the simulations.

The parameter $\theta$ is an arbitrary factor that sets the length of one timestep. We choose to use a calibration in which one timestep is five days and a year has $250$ trading days. This calibration seems reasonable considering several factors such as the average rate of return to fund investors, which would be $\sim8\%$ (for high leverage scenarios, not considering defaults) or the benchmark rate of return, which would be $15\%$.

The benchmark volatility $\sigma_{b}$ is set empirically to a very low volatility that is only reached in tranquil times. Alternatively it would be possible to refrain from using $\sigma_{b}$ and set $H_{\mathrm{min}}$ and $\Phi$ or $P_{\mathrm{max}}$ directly.

The parameter $S$ is a risk premium (spread) to a benchmark interest rate $i_{b}$, see above. We use a fixed spread $S$ for all fund managers of $0.75\%$. This low risk premium of approximately $1\%$ reflects the fact that loans are fully collateralized in the model.

\subsection{Returns and correlations} \label{returns} 
\begin{figure}
	\begin{center}
		\includegraphics[width=6.5cm]{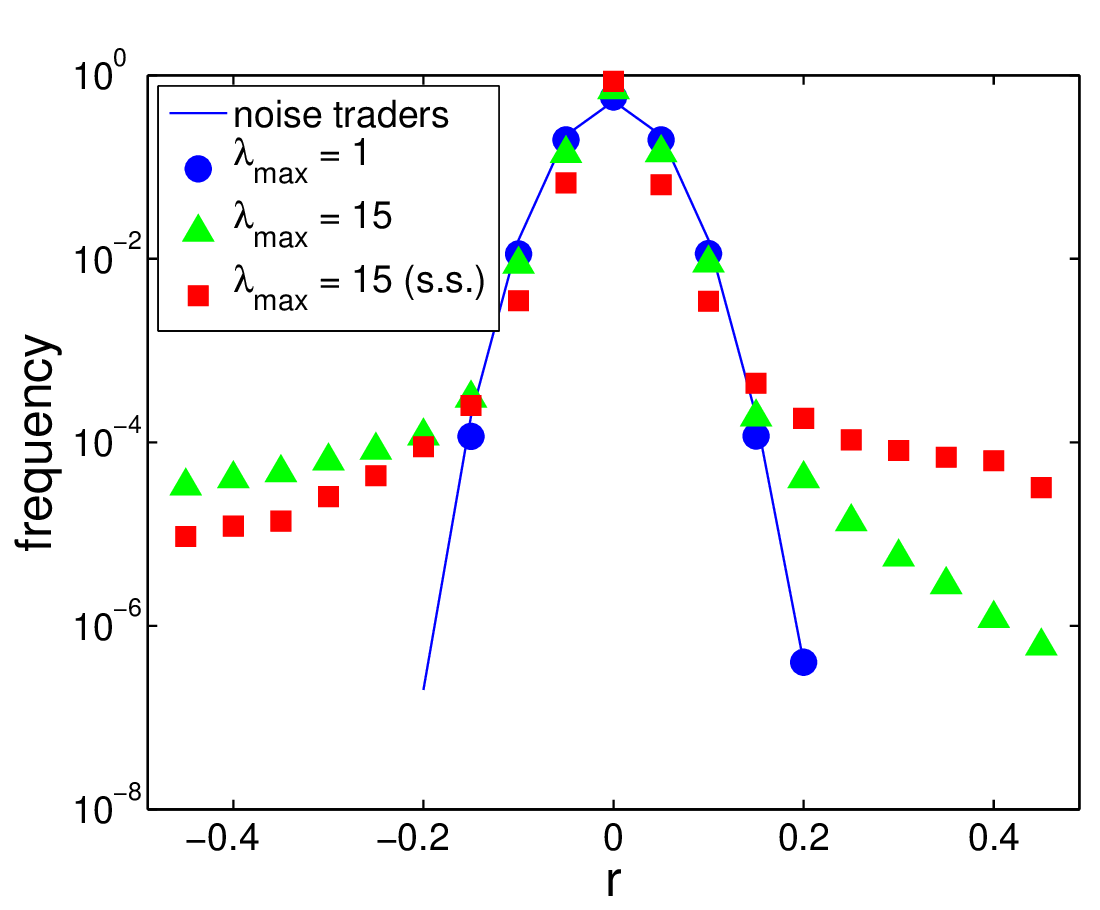} 
		\includegraphics[width=6.5cm]{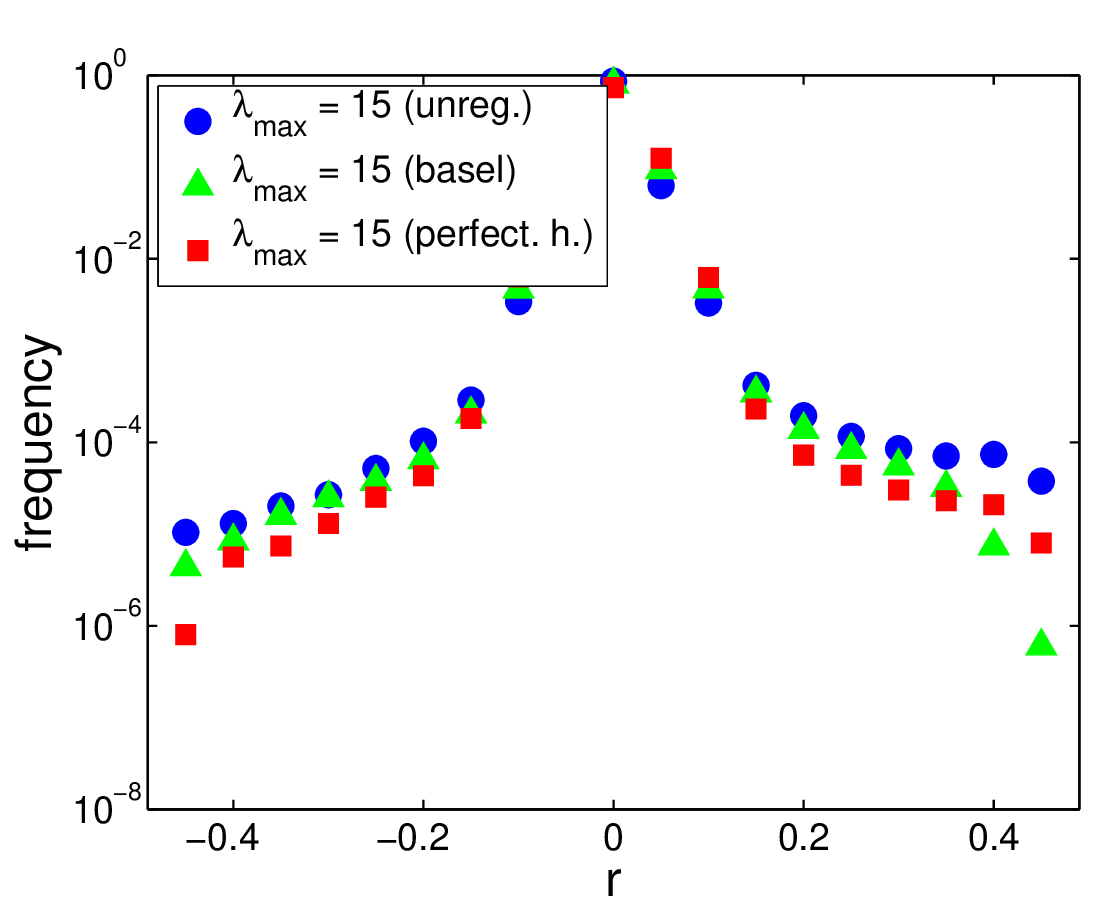} 
	\end{center}
	\caption[The distribution of $\log$-returns $r$]{The distribution of $\log$-returns $r$. (a) Return distributions of the baseline model (with parameters from subsection \ref{parameters}) in semi-log scale. The un-leveraged case (blue circles) practically matches the case with only noise traders (blue curve). When the maximum leverage is raised to $\lambda_{\mathrm{max}}=15$ (red squares) the distribution becomes more leptokurtic, and negative returns develop a fat tail. With short selling (demand equation \eqref{fundDemandshort}) the distribution becomes even thinner and tails turn fat on both sides. (b) Return distributions for the three regulation schemes at $\lambda_{\mathrm{max}}=15$. } \label{returnDist} 
\end{figure}

The statistical properties of asset price returns change considerably with increasing leverage. Figure~\ref{returnDist} shows the distribution of $\log$-returns $r(t)$ for four cases: (i) For the case of noise traders only, $\log$-returns are almost normally distributed. (ii) With un-leveraged fund managers, volatility is slightly reduced but $\log$-returns remain nearly normally distributed. (iii) When leverage is increased to $\lambda_{\mathrm{max}} = 15$ and no short selling is allowed, the distribution becomes thinner for small $r(t)$ but the negative returns develop fat tails. The asymmetry arises because with a short selling ban in place fund managers are only active when the asset is underpriced, i.e. when the mispricing $m(t) > 0$. Finally, when short selling is allowed (iv), the distribution becomes yet more concentrated in the center and fat tails develop on both sides. Due to higher risk involving short selling, the distribution becomes slightly asymmetric. This higher short selling risk arises because of the different risk profile of long and short positions. The potential losses from long positions are limited, since the price cannot go below zero. This is not the case for short positions, where the loss potential has no limit. Note that in all cases the autocorrelation of signed returns is very small, whereas the autocorrelation of absolute returns is strongly positive and decays slowly.

\subsection{Timeseries} \label{timeseries} 
\begin{figure}
	\begin{center}
		\includegraphics[width=14.0cm]{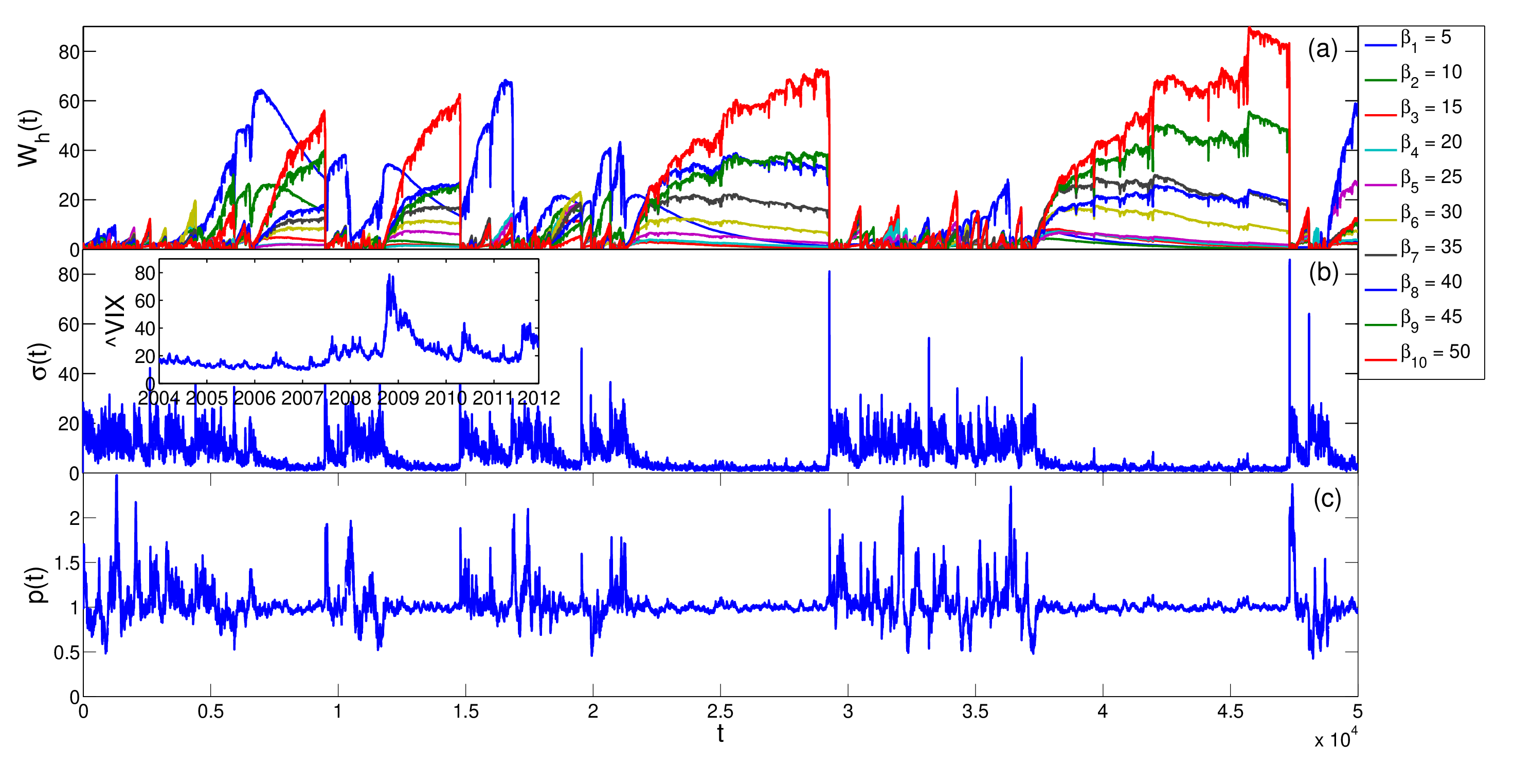} 
	\end{center}
	\caption[Timeseries for perfect-hedge scheme for $10$ fund managers.]{ Timeseries for perfect-hedge scheme for $10$ fund managers with $\beta_{h} = 5, 10, \ldots, 50$, for $\lambda_{\mathrm{max}}=15$. The simulation was done for the perfect-hedge scheme with maximum leverage equation \eqref{lambdamaxhedge} and wealth equations \eqref{fundWealthHedgeLong} and \eqref{fundWealthHedgeShort}. Simulation parameters are listed in subsection \ref{parameters}. (a) Wealth timeseries $W_{h}(t)$ of the fund managers. (b) Historical volatilities over a $\tau=10$ timestep window. For reference, in a market with noise traders only, the volatility is about $\sigma(t) \approx 17.5$ percent. For comparison the inset shows the VIX (Chicago Board Options Exchange Market Volatility Index), a measure of the implied volatility of S\&P 500 index options from 2004 to 2012. Our model reproduces the asymmetric profile in which volatility bursts are initiated by a rapid rise followed by a gradual fall. (c) Timeseries of the asset price $p(t)$.} \label{timeseriesfig} 
\end{figure}

The timeseries shown in figure \ref{timeseriesfig} are computed for the perfect-hedge scheme when short selling is allowed at $\lambda_{\mathrm{max}}=15$. Figure \ref{timeseriesfig}(a) shows the wealth $W_{h}(t)$ for all $10$ fund managers over time, figure \ref{timeseriesfig}(b) shows the historical asset volatility and figure \ref{timeseriesfig}(c) shows the timeseries of the asset price $p(t)$. The leverage cycle can be observed in this figure. Initially, all fund managers start with $W_h(0)=2$ and have a marginal influence on the market. As they gain more wealth their market impact increases, mispricings are damped and the asset volatility decreases. The decrease in volatility results in lower borrowing cost for the fund managers. Lower borrowing cost allows them to use a higher leverage, which further lowers volatility, resulting in even lower borrowing costs etc.. This leads to seemingly stable market conditions with low volatility and low borrowing costs for fund managers; such stable periods can persist for a long time. Occasionally the stable periods are interrupted by crashes. These are triggered by small fluctuations of the noise trader demand. As shown in \cite{thurner09}, if one or more of the funds is at its leverage limit, a downward fluctuation in price causes the leverage to rise. This triggers a margin call, which forces the fund to sell into a falling market, amplifying the downward fluctuation. This may cause other funds to sell, driving prices even further down, to cause a crash. Such crashes can trigger price drops as large as 50\%, which cause the more highly leveraged fund managers to default. After a crash noise traders dominate and volatility is high. Fund managers are reintroduced; as their wealth grows volatility drops, and the leverage cycle starts again. In situations where a crash wipes out all but the least aggressive fund managers, as happens around $t=29,000$ and at about $t=47,000$, the surviving less aggressive fund managers become dominant for extended periods of time.

\subsection{Volatility profile}

One of the interesting aspects of our model is that it reproduces the asymmetric profile of volatility bursts. The inset of figure \ref{timeseriesfig}(b) shows the VIX (Chicago Board Options Exchange Market Volatility Index), which is a measure of the implied volatility of S\&P 500 index options, from 2004 to 2012. As can seen by eye in this figure, in our model, as in the real data, volatility bursts rise rapidly and damp out gradually, forming a very characteristic asymmetric peak. 
\begin{figure}
	\begin{center}
		\includegraphics[width=14.0cm]{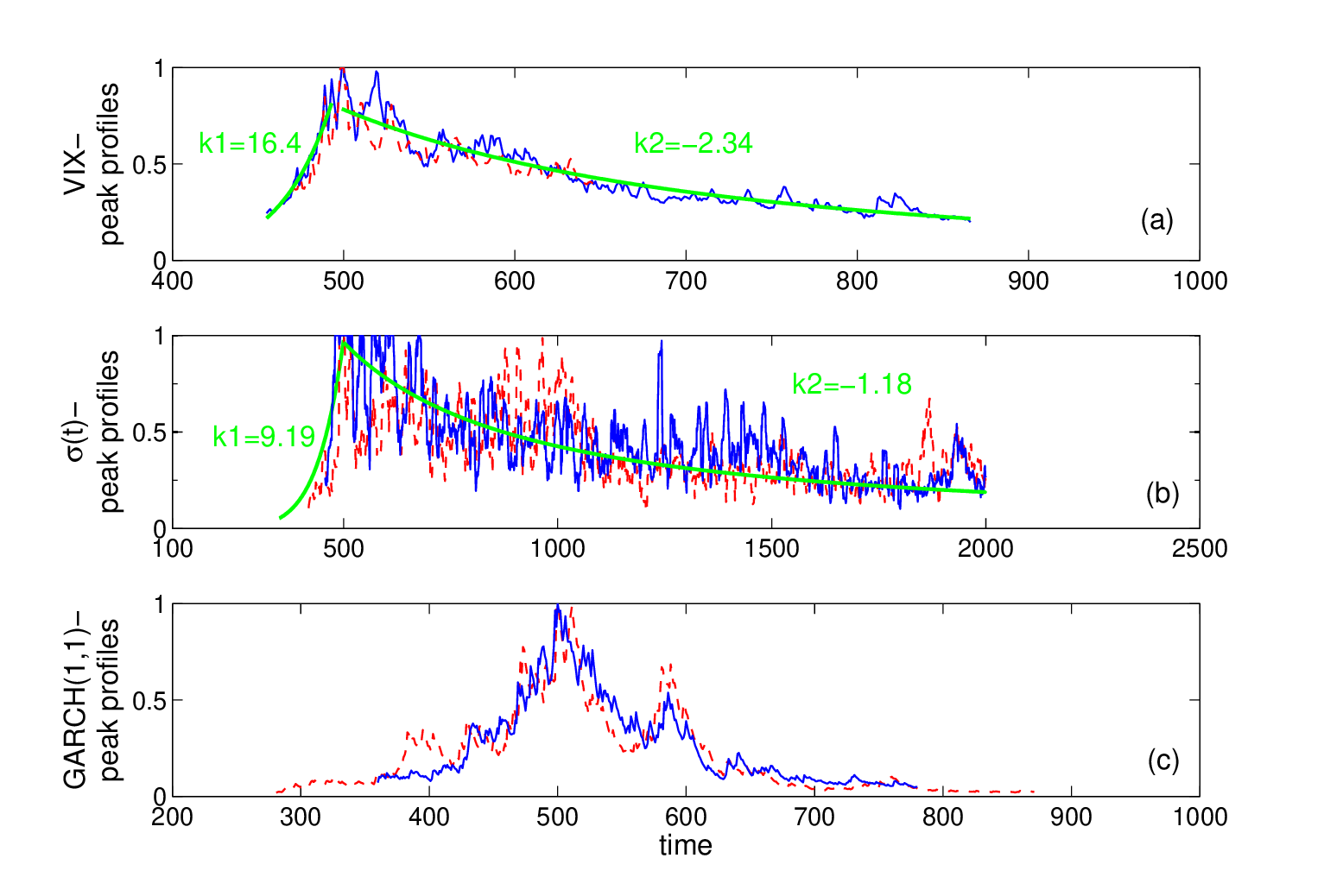} 
	\end{center}
	\caption[Peak profiles of the VIX (a), our model (b) and a sample path simulated with a GARCH(1,1) model (c).]{Peak behavior of the VIX (a), our model (b) and a GARCH(1,1) model (c). We find that the rise and fall of volatility during a peak follows roughly a power law of the form $\sigma(t) \appropto t^k$. The fitted values for the rise of the VIX are $k_1\sim16.39$ and for our model $k_1\sim9.19$. The VIX gradually falls with $k_2\sim-2.34$ and for our model with $k_2\sim-1.18$. GARCH models do not exhibit power-law behavior, and $k_1$ and $k_2$ can not be fitted. In each case we show two characteristic time series in blue (solid line) and red (dashed line), normalized to have the same peak height. The time axis is shifted so that peaks occur at $t=500$. Otherwise, the time axis is not normalized or gauged to real time.} \label{peakprofile} 
\end{figure}

To see this more quantitatively in Figure \ref{peakprofile} we compare the peak behaviors of the VIX, our model and a GARCH(1,1) model. Two example time series around a peak are shown for each. The data is normalized so that the peaks are all of the same height. The time axis is shifted so that peaks occur at $t=500$. Otherwise, the time axis is not normalized or gauged to real time. We fit the rise and decay around the peaks as a power law of the form $\sigma(t) \appropto t^k$, characterizing the rise by $k=k_1$ and the decay by $k=k_2$, and take the average for the largest peaks in the time series, both from the VIX and from our model. To determine which peaks to fit, we selected local maxima that are at least $(\max(\text{VIX})-\min(\text{VIX}))/4)$ or $(\max(\sigma(t))-\min(\sigma(t)))/4)$ above the surrounding data. The fitted values for the rise of the VIX are $k_1\sim16.39$ and for our model $k_1\sim9.19$. The VIX gradually falls with $k_2\sim-2.34$ and for our model with $k_2\sim-1.18$. GARCH models do not exhibit power-law behavior and $k_1$ and $k_2$ cannot be fitted. Note that while the form of the peaks is the same for the VIX and our model, the magnitude is not; also the VIX follows the profile with less noise than our model does. A possible reason for this is that the VIX shows the implied volatility of S\&P 500 index, while our model shows the volatility of a single asset.

As a further statistical comparison we calculated the standardized moments skewness and kurtosis. The skewness of the VIX ($\sim2.02$) is virtually identical to our model ($\sim2.01$), however the kurtosis of the VIX ($\sim8.13$) is lower compared to our model ($\sim14.59$). Note that the VIX shows even in tranquil times quite substantial volatility in comparison to our model, where extended periods with almost no volatility can be observed. The reason for this is that in times of high market share of the fund managers, the effect on the asset price of the noise traders is negligible. All fund managers have the same perceived fundamental value $V$ and thus stabilize the asset price to a point where almost no volatility can be observed. With a different fundamental value $V_h$ for every fund manager the behavior would become less synchronized and therefore, even in tranquil times, volatility would remain at a substantial level. We have explicitly checked this in a series of experiments.

\subsection{Comparison of regulatory schemes} \label{impacts} 
\begin{figure}
	\begin{center}
		\includegraphics[width=6.5cm]{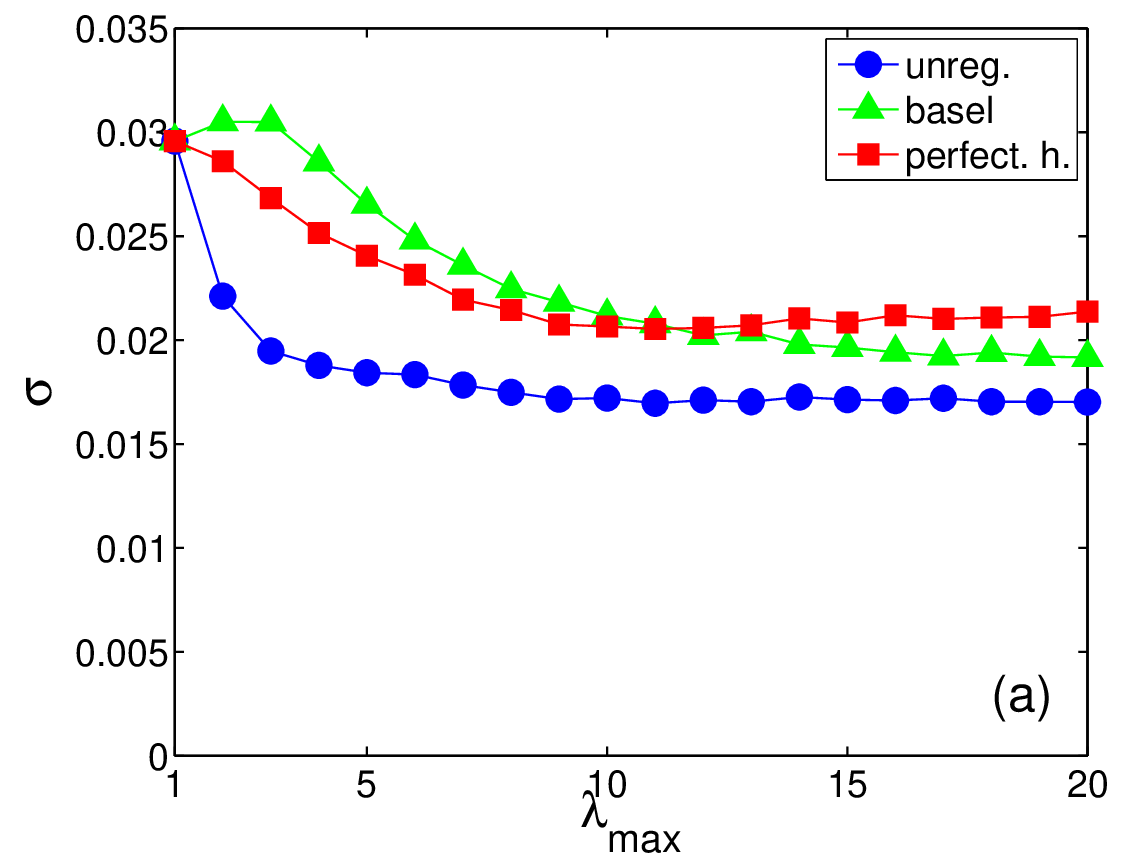} 
		\includegraphics[width=6.5cm]{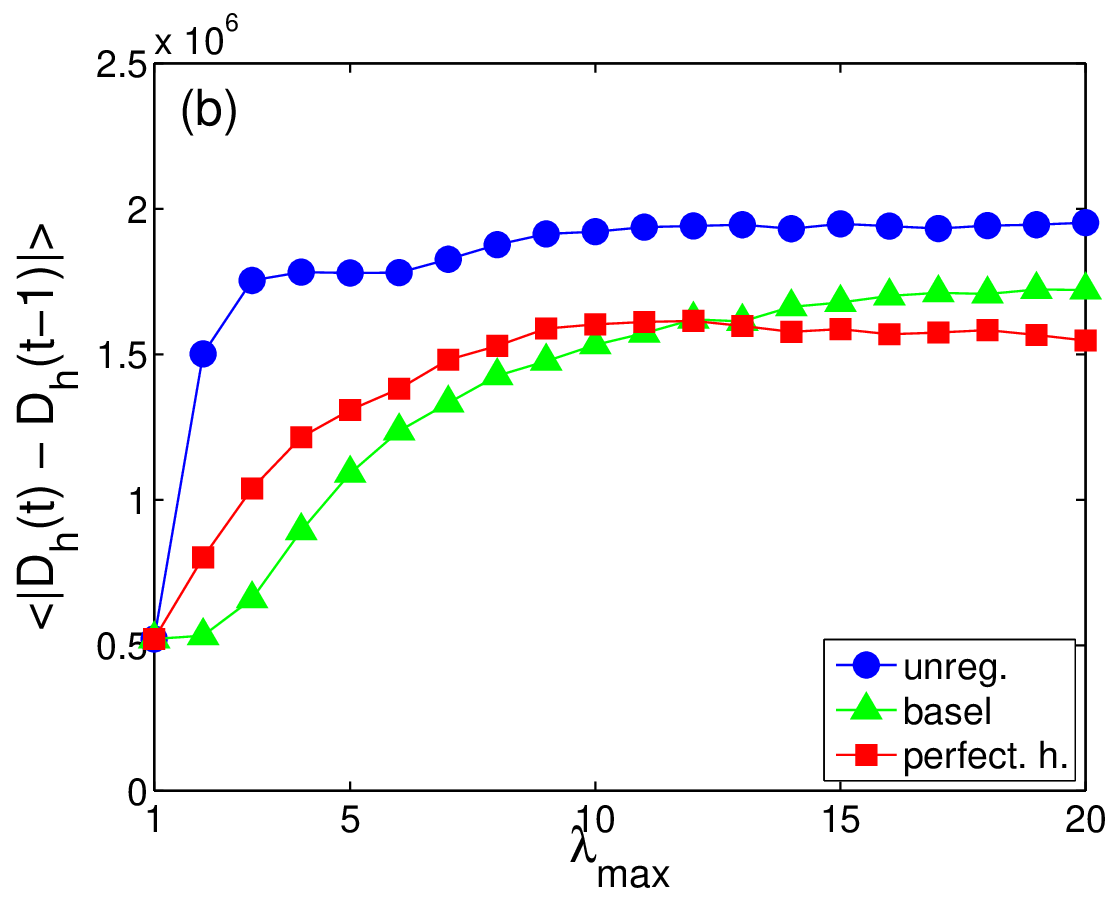} 
		\includegraphics[width=6.5cm]{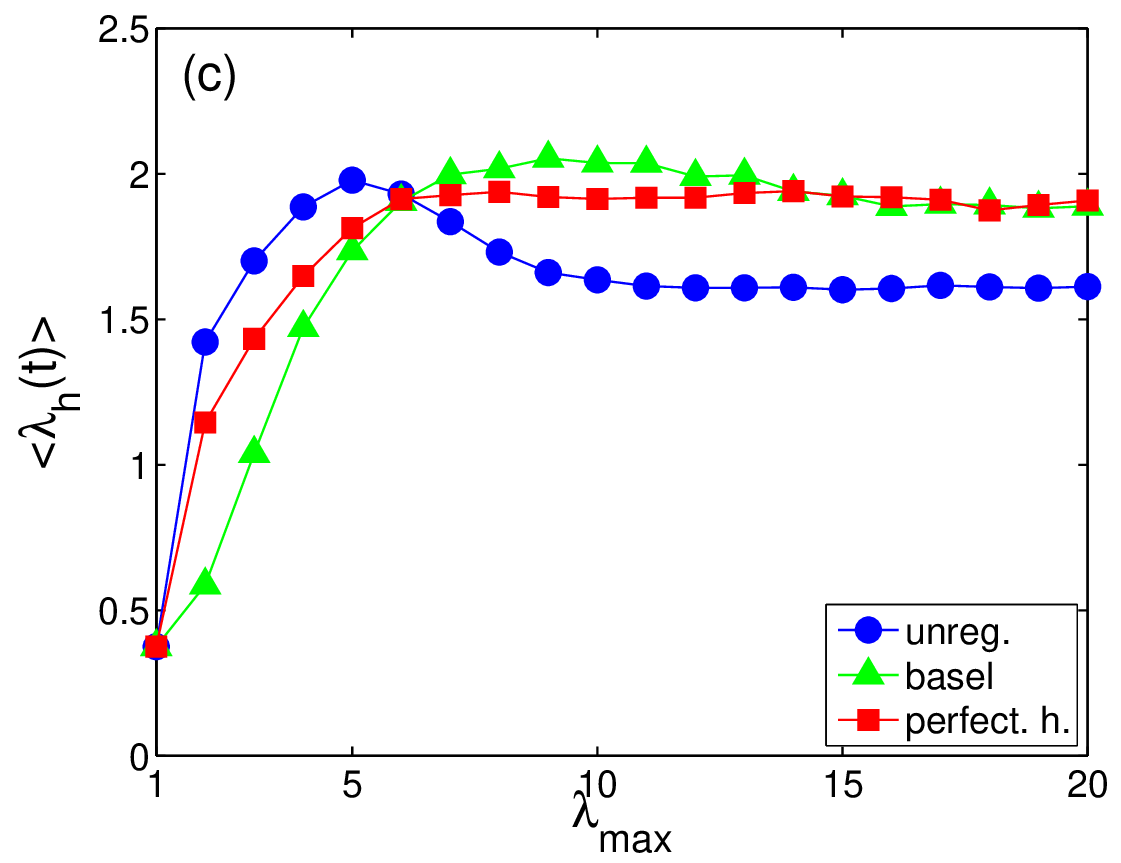} 
		\includegraphics[width=6.5cm]{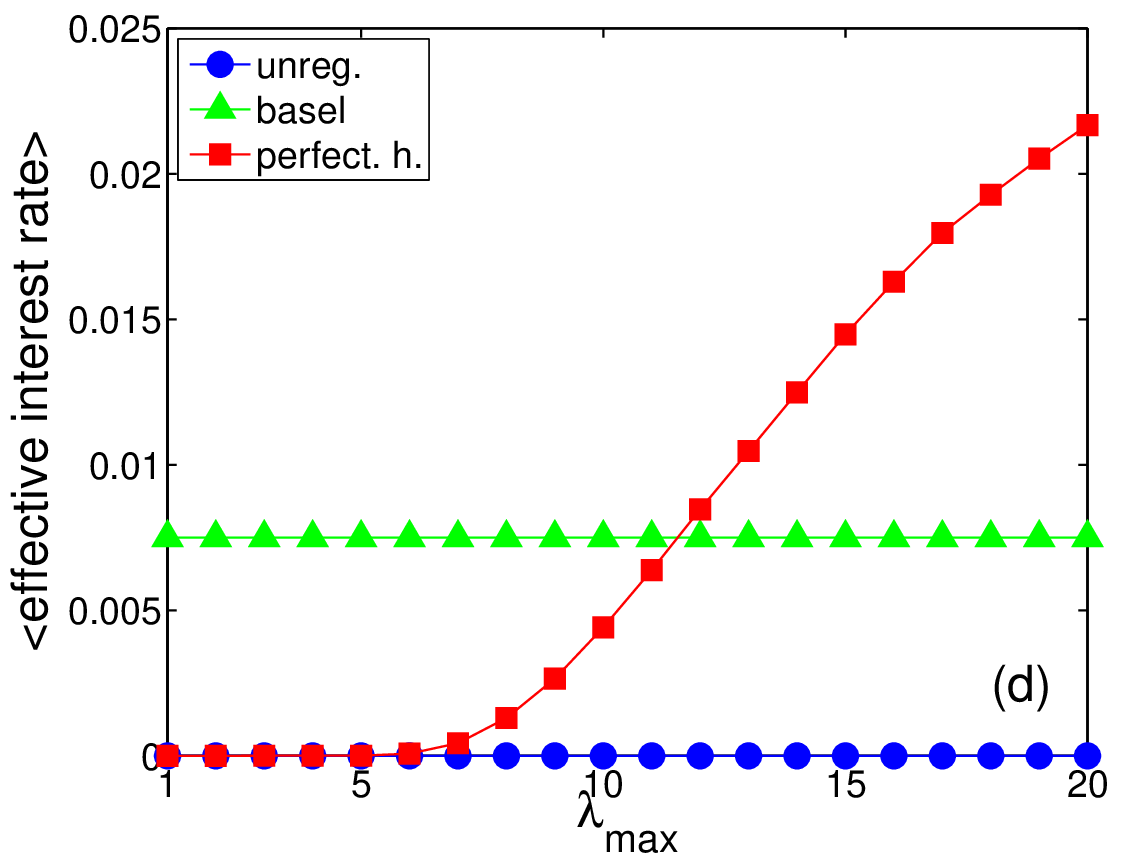} 
	\end{center}
	\caption[Impacts of regulatory measures on market indicators as the maximum leverage varies.]{Impacts of regulatory measures on market indicators as $\lambda_{\rm max}$ varies. (a) Volatility of the underlying asset. (b) Market volume of the underlying asset, measured by the average amount of shares traded by a fund manager per timestep. (c) Average effective leverage of fund managers (d) Average annualized effective interest rate. For all simulations we used 10 fund managers with $\beta_{h} = 5, 10, \ldots,50$ over $5\times10^4$ timesteps. Simulation parameters listed in subsection \ref{parameters}, short selling was allowed. For all indicators it is assumed that one timestep takes five days and a year has 250 trading days. Blue, green, and red curves indicate the unregulated, the Basle II, and the perfect-hedge scheme, respectively. Standard deviations computed over $100$ independent runs are below symbol size. } \label{marketimpact} 
\end{figure}
\begin{figure}
	\begin{center}
		\includegraphics[width=6.5cm]{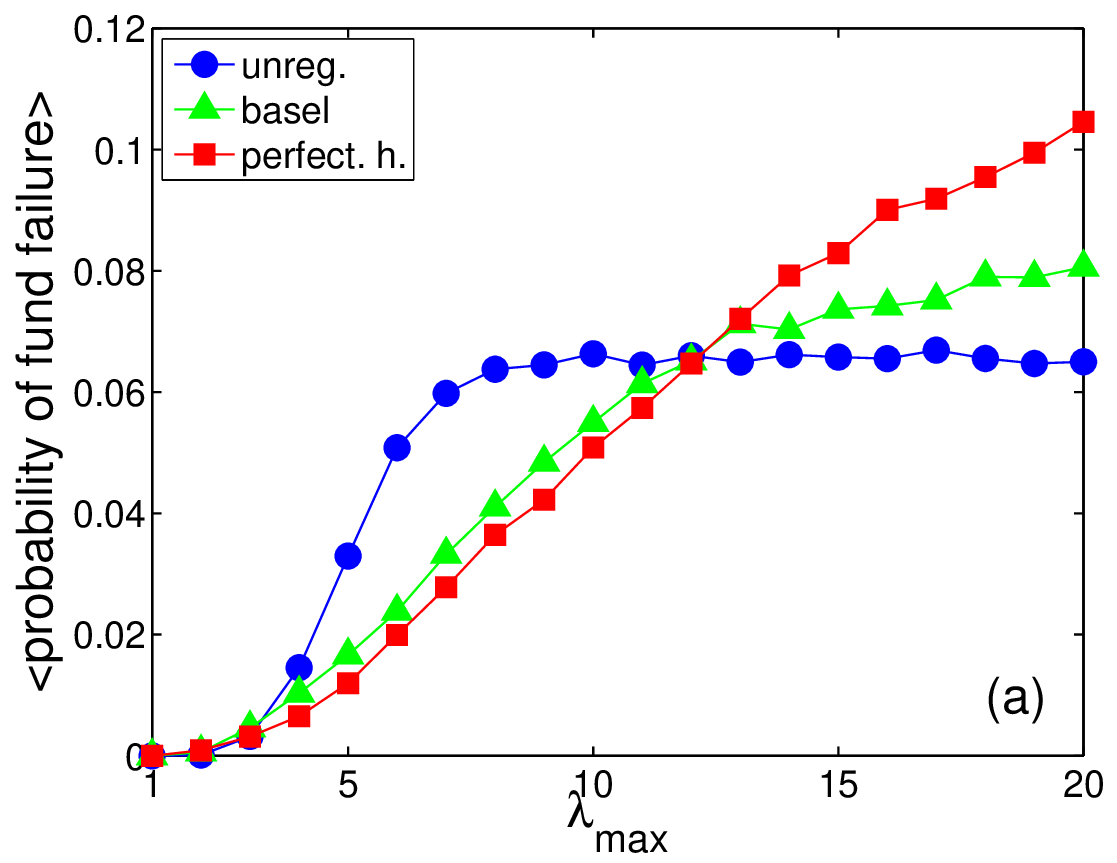} 
		\includegraphics[width=6.5cm]{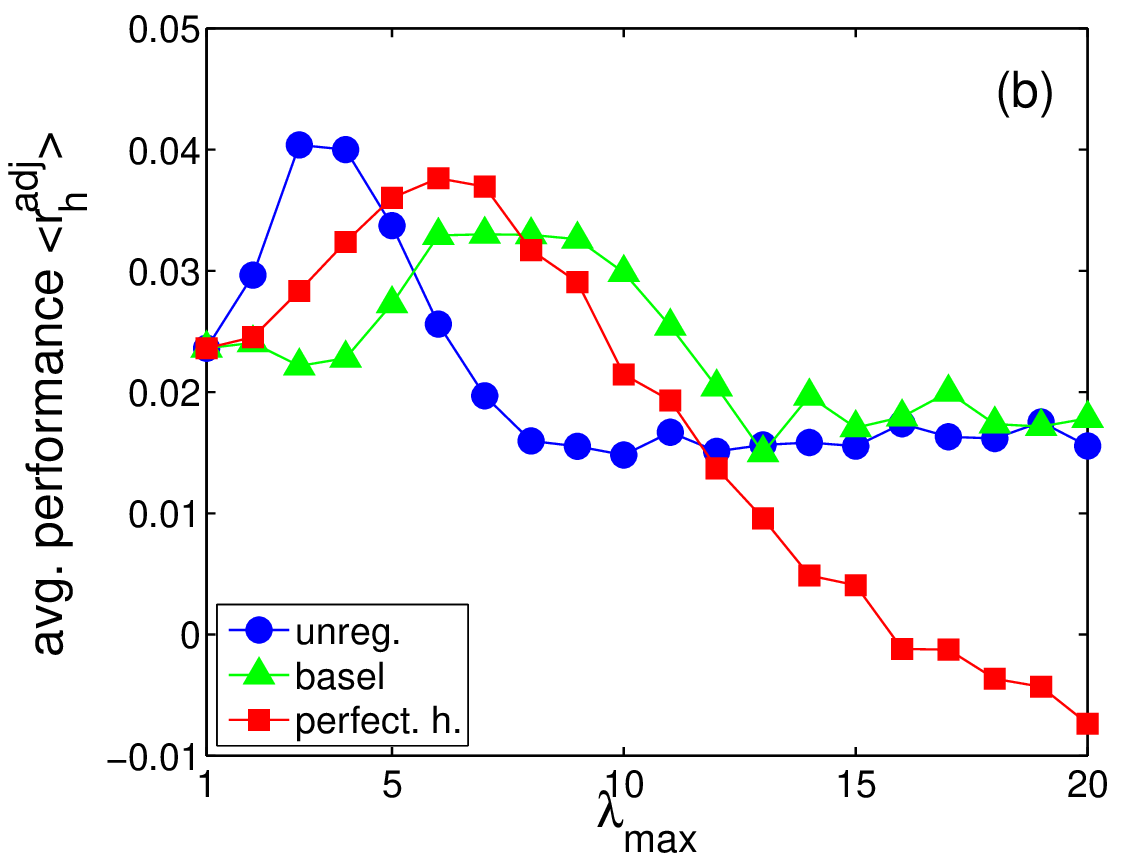} 
		\includegraphics[width=6.5cm]{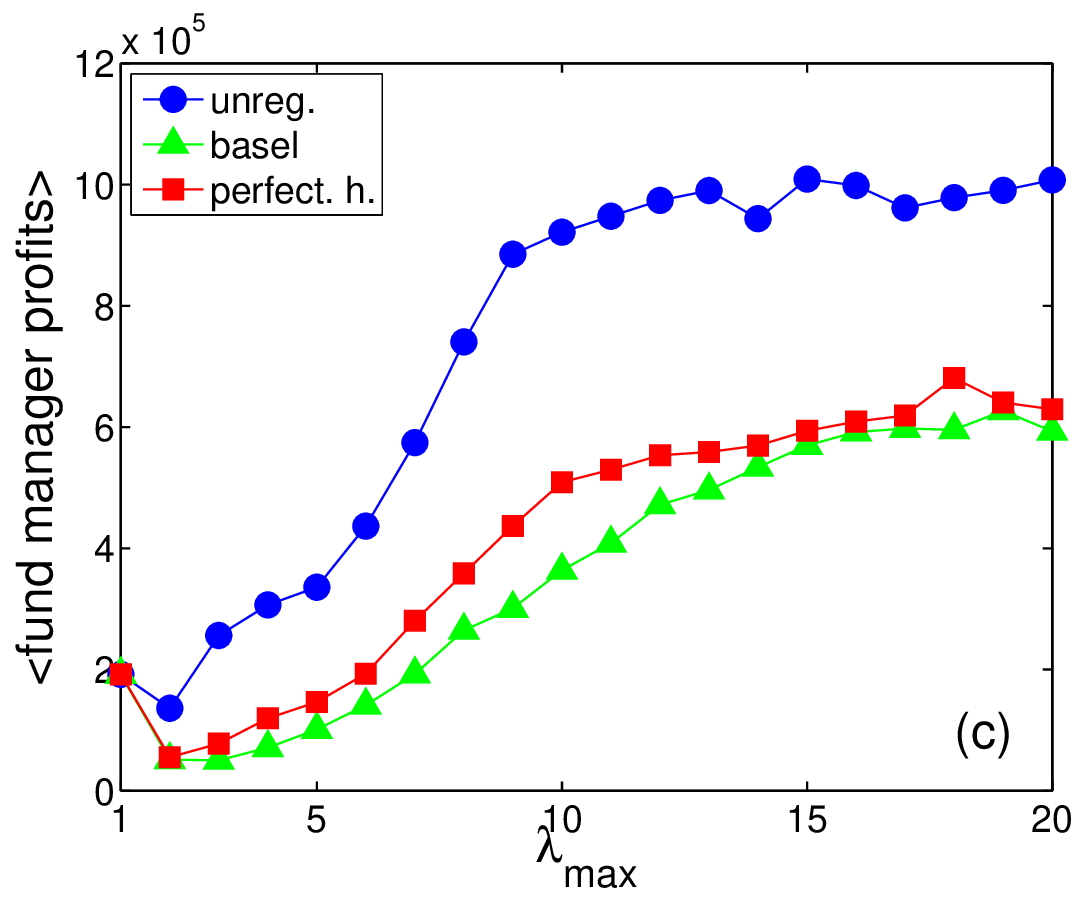} 
		\includegraphics[width=6.5cm]{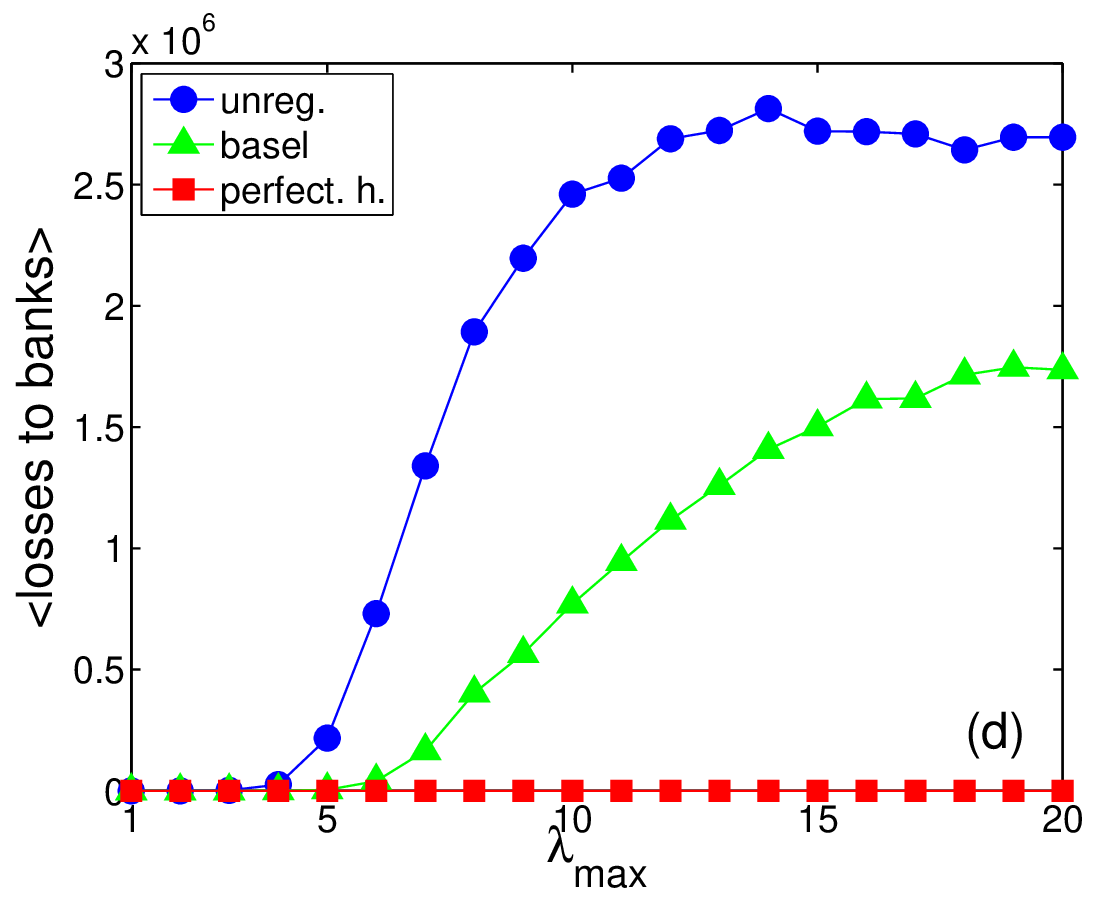} \caption[Impacts of regulatory measures on indicators for the most aggressive fund manager.]{Impacts of regulatory measures on performance indicators for the most aggressive fund manager with $\beta_{h} = 50$ as $\lambda_{\rm max}$ varies. (a) Average annual probability of fund failure. (b) Average annual adjusted rate of return $r^{adj}$ to fund investors. (c) Average annual profits to the fund managers. (d) Annual capital shortfall of banks. For all simulations we used the same setup as in the previous figure. Standard deviations computed over $100$ independent runs are about $0.01$, $0.01$, $2.7\times10^5$ and $5.4\times10^5$ for (a), (b), (c) and (d), respectively. } \label{fundimpact} 
	\end{center}
\end{figure}

In the following we illustrate the impacts of the three regulatory schemes: 
\begin{enumerate}
	\item The unregulated baseline scheme from section \ref{baselinemodel} with fixed maximum leverage $\lambda_{\mathrm{max}}$, corresponding to demand equation \eqref{fundDemandshort}. 
	\item The Basle II scheme, which has the same demand equation but with $\lambda_{\rm max}$ replaced by $\lambda^{\mathrm{adapt}}_{\mathrm{max}}(t)$ of \eqref{lambdamaxbasel} and the wealth update given by \eqref{fundWealthSpreadLong} and \eqref{fundWealthSpreadShort}. 
	\item The perfect-hedge scheme, where $\lambda^{\mathrm{adapt}}_{\mathrm{max}}(t)$ is given by \eqref{lambdamaxhedge} and the wealth update is \eqref{fundWealthHedgeLong} and \eqref{fundWealthHedgeShort}. 
\end{enumerate}
For all simulations we used the parameters listed in subsection \ref{parameters} and enabled short selling. $\lambda_{max}$ was varied from $1$ to $20$. For each parameter set we compute $5\times10^4$ timesteps and average over 100 independent runs. The following indicators are annualized using a calibration in which one timestep is five days and a year has $250$ trading days. This calibration seems reasonable considering several factors such as the average rate of return to fund investors, which would be $\sim8\%$ (for high leverage scenarios, not considering defaults) or the benchmark rate of return, which would be $15\%$.

Figure \ref{marketimpact} provides a panel of results that show how market characteristics such as (a) price volatility, (b) trading volume, (c) average leverage and (d) the effective interest rate are affected by the varying the maximum leverage parameter $\lambda_{\mathrm{max}}$ under each of the three regulatory schemes. For the unregulated case the volatility drops monotonically as $\lambda_{\mathrm{max}}$ increases, dropping by almost a factor of two for $1 < \lambda_{\mathrm{max}} < 10$ and then reaching a plateau. The reason the volatility drops is that the presence of value investors ordinarily damps volatility, since they buy when prices fall and sell when prices rise. Larger leverage means more trading and it also means higher profits for a given level of trading, which means investors place more money in the funds and they have more market power. The resulting increase in trading is evident in Figure \ref{marketimpact} (b): Between $\lambda_{\mathrm{max}} = 1$ and $\lambda_{\mathrm{max}} = 3$ the trading volume increases by more than a factor of three, and then more or less levels off. 

Volatility is not significantly influenced by failures of fund managers. Bankruptcies of the most aggressive fund manager occur on average every $800$ time steps for $\lambda_{\mathrm{max}} > 7$. Therefore the subsequent price drops or jumps do not occur frequently enough to substantially influence volatility.

The behavior of the volatility under the Basle II scheme is more complicated, as is evident in Figure \ref{marketimpact} (a). The volatility initially rises, reaching a maximum at $\lambda_{\mathrm{max}} \approx 2$ and then falls, although never to the low level of the unregulated case. 

The explanation for the initial rise in volatility comes from the fact that for $\lambda_{\rm max}=1$ there is no short selling possible, see Eq. (\ref{fundDemandshort}). In this case there are long-only funds only. As $\lambda_{\rm max}=2$, funds now also engage in short selling and we experience an effective regime-shift from a long-only to a long-and-short model. Note that the average leverage for the Basle II regime (Fig. \ref{marketimpact} (c)) is below $1$, meaning that there are often cases where there is no leverage. In those situations where a leverage reduction from $\lambda> 1$ to $\lambda=1$, fund managers are forced out of short positions leading to do extra trading activity (that does not exist for the $\lambda_{\rm max}=1$ case) that slightly raises the volatility. 

The behavior of the perfect hedge scheme is also more complicated; it initially falls until $\lambda_{\rm max} \sim 10$, but then rises again. 

The reason for the increase in volatility for large $\lambda_{\rm max}$ under the perfect hedge scheme is that for high maximum leverage the effective borrowing costs are large, as illustrated in Figure \ref{marketimpact} (d). This is also the reason why in this scheme the trading volume drops for $\lambda_{\rm max} > 10$. (See equations~\eqref{fundInterestSpread} and \eqref{fundInterestHedgeLong}). For the unregulated case the borrowing costs are zero (since for convenience we set the base interest rate to zero), whereas under the Basle II scheme they are fixed. In the perfect hedge scheme the costs are near zero for small leverage and then grow sharply starting at about $\lambda_{\rm max} \approx 5$, and exceed the costs for the Basle II scheme at about $\lambda_{\rm max} \approx 12$. Thus the perfect hedging scheme is cheaper than the Basle II scheme for low leverage and more costly for high leverage.

Figure \ref{marketimpact} (c) shows the average effective leverage of fund managers, defined as
\[ \langle \lambda_h(t) \rangle = \frac{1}{HT}\sum^H_{h=1}\sum^T_{t=1}\lambda_{h}(t). \]
For the unregulated case, as expected, the average leverage initially increases as $\lambda_{\rm max}$ increases. It rises from $\langle \lambda_h(t) \rangle \approx 0.4$ when $\lambda_{\rm max} = 1$ to a little less than two when $\lambda_{\rm max} \approx 5$. Surprisingly, however, it then decreases until about $\lambda_{\rm max} \approx 10$, settling into a plateau for $\lambda_{\rm max} > 10$ with $\langle \lambda_h(t) \rangle \approx 1.5$. The reason for the decrease is the drop in volatility, which means that there are fewer mis-pricings and less opportunities to use leverage. The Basle II and perfect hedge schemes behave similarly, except that they drop very little after their peaks, which are reached at larger $\lambda_{\rm max}$, and their plateau leverage is higher, with $\langle \lambda_h(t) \rangle \approx 2$.

Figure \ref{fundimpact} shows a variety of diagnostics about market performance, including (a) the probability of fund default, (b) return to investors, (c) profits for fund managers and (d) capital shortfall of banks. 

Figure \ref{fundimpact} (a) shows the average annual probability of default for the most aggressive fund manager. In the unregulated scheme the annual probability of default initially grows rapidly, but then reaches a plateau at $\lambda_{\rm max} \approx 8$. The reason for this is that under the unregulated scheme the use of leverage is automatically self-limiting due to dropping volatility, as demonstrated in Figure~\ref{marketimpact}(c). For the Basle II and perfect hedge schemes the initial rise in defaults is slower, but unlike the unregulated case the default rate never plateaus, and default exceeds the unregulated scheme for roughly $\lambda_{\rm max} > 11$. The reason that there are more defaults in both regulated schemes is that the maximum leverage is adjusted dynamically. If the maximum leverage is suddenly decreased when funds are at their maximum leverage, they are forced to sell {\it en masse}, and the resulting market impact can trigger a crash\footnote{ 

For $\lambda_{\rm max} <10$ the perfect-hedge scheme performs a bit better than the Basle II scheme because of the stronger limit to lending based on historical volatility illustrated in figure \ref{lambdamax}.}.

In figure \ref{fundimpact}(b) we show the average adjusted rate of return $r^{adj}$ to fund investors for the most aggressive fund manager. The return to fund investors is influenced mainly by three factors: effective leverage, mis-pricing opportunities and probability of fund default. In each of the three regulatory schemes the investor returns behave similarly, initially increasing with $\lambda_{\rm max}$, reaching a peak, and then decreasing. Under the Basle II and perfect hedge schemes the peaks come later: For the unregulated scheme the peak is at $\lambda_{\rm max} \approx 4$, for the perfect hedge scheme it is at $\lambda_{\rm max} \approx 6$, and for Basle II it is at roughly $\lambda_{\rm max} \approx 8$. The primary reason there is a peak is the rising default rate\footnote{

If funds that have gone bankrupt are excluded the performance reaches a maximum at $\lambda_{\rm max} \sim7$.}.

Another important factor is decreased volatility, and hence fewer mispricings and lower effective leverage. For large $\lambda_{\rm max}$ the perfect hedge case behaves increasingly poorly due to sky-rocketing borrowing costs (see Figure \ref{marketimpact}(d)). 

Figure \ref{fundimpact}(c) shows the fund managers' profits. We assume a hypothetical $2\%$ fixed fee for the assets under management and a $20\%$ performance fee. Profits to funds are consistently higher in the unregulated scheme than in either of the alternatives. The discrepancy is exaggerated relative to investors' profits due to the fact that under the unregulated scheme the assets under management by the funds are much larger. It is perhaps not surprising that fund managers clearly prefer less regulation. Surprisingly, as $\lambda_{\rm max}$ increases from one to two, fund managers' profits initially drop; for the unregulated case it reaches a minimum about 30\% less than the unleveraged case, which for the regulated schemes is almost 75\%. The origin of this problem is the same as mentioned above, that for $\lambda_{\rm max}=1$ it is not possible to short sell, see Eq. (\ref{fundDemandshort}). In this case we have long-only funds. As soon as we bring leverage to $\lambda_{\rm max}=2$, funds engage in short selling and we have a shift from a long-only to a long-and-short model. We have explicitly checked this issue by tests, where we make the demand function for short selling Eq. (\ref{fundDemandshort}) fully symmetric to the long-part Eq. (\ref{fundDemand}), by removing the $1$ in Eq. (\ref{fundDemandshort}). In this case no more drop in the managers' profits from $\lambda_{\rm max}=1$ to $\lambda_{\rm max}=2$ occurs. Note that this symmetric implementation is of course not the correct one. Profits are consistently higher for the perfect hedge scheme than they are in the Basle II scheme. 

Finally, in Figure \ref{fundimpact}(d) we show the capital shortfall of banks. As expected the shortfall is the largest for the unregulated case, and for the perfect hedge case it is zero by definition.

\section{Discussion} \label{discussion}

We studied an agent based model of a financial market where investors have different degrees of information about the fundamental value of financial assets. Fund managers leverage their investments by borrowing from banks. These speculative investments based on credit introduce a systemic risk component to the system. When the collateral backing the involved loans are composed of the same financial assets themselves, the model reproduces the various stages of the leverage cycle, \cite{geanakoplos97, geanakoplos03,fostel08, geanakoplos10}, and the systemic prerequisites leading up to crashes can be studied in detail, \cite{thurner09}. In this work we have studied the role of Basle II-type regulations, which define the conditions under which banks can provide leverage. These regulations basically require Banks to apply haircuts to the collateral accepted, and to take spreads. This raises the capital costs for the leverage takers, so that they take less effective leverage than they would in the unregulated case. We find that this is indeed the case for situations of low leverage in the system; regulation makes the system more secure. On the other hand regulation reduces the market share of fund managers in general, with the consequence that they take less volatility from the markets than they would in the unregulated case. In this sense there is no optimal level of leverage or optimal level of regulation. Regulation makes the markets more secure in the sense that it reduces the frequency of large price jumps, but renders markets more volatile in general.

The situation changes drastically when the general level of leverage is high in the system, which is the case for phases of low-volatility. When the leverage is greater than about 10, investors in the regulated schemes take more effective leverage than they would in an unregulated world. This is easy to understand: in the unregulated situation the fund managers become bigger (in the model up to twice as big in terms of assets under management) and hence reduce volatility more effectively. In this scenario the regulated system becomes less stable than the unregulated one, which is for example reflected in a higher default rate of regulated investors than unregulated ones. Also the regulated system does not manage to reduce the frequency of crashes. In terms of capital shortfall of banks, the regulated system is superior to the unregulated one for all leverage levels in the system.

We next designed a hypothetical regulation system, where banks require their investors to hedge against the risk of loss of collateral value. Under the assumption that the writers of options never default, by construction the capital shortfall of banks is zero. This regulation scheme tests the effect of making the capital costs (effective interest rate) for the fund managers dependent on the actual leverage taken; capital for higher leverage (and thus more systemic risk) is more expensive. This is not the case for Basle II-type regulation, where the interest rate is independent of leverage. 

Even under the perfect hedging scheme the system does not get much safer on a systemic level. The effects are qualitatively and quantitatively similar to the Basle II-type scenario. However, in terms of volatility reduction the perfect-hedge scheme is shown to be superior to the Basle II scheme.

We have thus demonstrated that even under the assumption of a perfect-hedge scheme, systemic risk originates from a source that is not addressed by present regulation mechanisms. Systemic risk arises due to a synchronization of the agents' behavior in times of high leverage. Such synchronized behavior can become worse under regulation. This is because a lowering of the maximum leverage at a point where some of the funds are fully leveraged can cause a wave of selling, driving prices down and triggering even more selling, and in some cases leading to a crash.

In this paper we do not propose a solution to the problem of credit regulation. However, our belief is that a key element of any such regulation should be greater transparency. As argued in \cite{thurner11}, one can imagine a system where every credit provider has to disclose the amount of its loans and the identity of the borrower on their homepage, and borrowers would have to disclose their leverage. In this way there would develop a transparent credit market with emerging lending rates, depending on the riskiness of the creditor and the borrower. Consequences of such schemes are presently under investigation.

\section*{Acknowledgements} Funding was provided by EU FP7 project CRISIS, agreement no. 288501, and EU FP7 project MULTIPLEX agreement no. 317532.

\end{document}